\documentclass[apj,numberedappendix]{emulateapj}
\usepackage{amsmath}
\usepackage{hyperref}

\usepackage[english]{babel}
\usepackage{graphicx}
\usepackage{latexsym}
\usepackage{amsmath,amssymb}
\usepackage{natbib}
\usepackage{dcolumn}
\usepackage{bm}
\usepackage{calrsfs}
\usepackage{ulem}
\usepackage[usenames]{color}
\usepackage{xspace}

\newcommand{\na}{New Astr.}

\newcommand{\nn}{\mbox{} \nonumber \\ \mbox{} }
\newcommand{\ba}{\begin{eqnarray}}
\newcommand{\ea}{\end{eqnarray}}

\newcommand{\E}{{\bf E}}
\newcommand{\B}{{\bf B}}

\newcommand{\Bf}{{magnetic field}}

\newcommand{\Bfs}{{magnetic fields}}
\newcommand{\NS}{neutron star}

\newcommand{\NSs}{{neutron stars}}
\newcommand{\Lf}{{Lorentz factor}}
\newcommand{\LC}{light cylinder}

\newcommand\eg{\textit{e.g.}}

\def\be{\begin{equation}}
\def\ee{\end{equation}}

\begin{document}

\title{GRB~170817A Associated with GW170817: Multi-frequency Observations and Modeling of Prompt Gamma-ray Emission}

\author{A.~S.~Pozanenko$^{1,2,3}$,
M.~V.~Barkov$^{4,5}$,
 P.~Yu.~Minaev$^{1}$,
 A.~A.~Volnova$^{1}$,
 E.~D.~Mazaeva$^{1}$,
 A.~S.~Moskvitin$^{6}$,
 M.~A.~Krugov$^{7}$,
 V.~A.~Samodurov$^{2,8}$,
V.~M.~Loznikov$^{1}$,
and
M.~Lyutikov$^{4}$
}

\affil{
$^{1}$Space Research Institute, 84/32 Profsoyuznaya Street, Moscow
117997, Russia \\
$^{2}$National Research University Higher School of Economics, Myasnitskaya 20, 101000,  Moscow, Russia \\
$^{3}$National Research Nuclear University MEPhI (Moscow Engineering Physics Institute), Kashirskoe shosse, 31, 115409 Moscow, Russia \\
$^{4}$Department of Physics and Astronomy, Purdue University, 525 Northwestern Avenue, West Lafayette, IN
47907-2036, USA\\
$^{5}$Astrophysical Big Bang Laboratory, RIKEN, 2-1 Hirosawa, Wako, Saitama 351-0198, Japan\\
$^{6}$Special Astrophysical Observatory of Russian Academy of
Sciences, Nizhniy Arkhyz, 369167, Russia\\
$^{7}$Fesenkov Astrophysical Institute, Almaty, 050020, Kazakhstan\\
$^{8}$Pushchino Radio Astronomy Observatory ASC LPI, Pushchino, Russia\\
}

\begin{abstract}

We present our observations of electromagnetic transients associated with GW170817/GRB~170817A using optical telescopes of Chilescope observatory and Big Scanning Antenna (BSA) of Pushchino Radio Astronomy Observatory  at 110~MHz. The Chilescope observatory detected an optical transient of $\sim19^m$ on the third day in the outskirts of the galaxy NGC 4993; we continued observations following its rapid decrease. We put an upper limit of $1.5\times10^{4}$~Jy on any radio source with a duration of 10--60 s which may be associated with GW170817/GRB~170817A. The prompt gamma-ray emission consists of two distinctive components -  a hard short pulse delayed by $\sim2$ seconds with respect to the  LIGO signal and softer
thermal pulse with $T\sim 10 $ keV lasting for another $\sim2$ seconds.  The appearance of a thermal component at
the end of the burst is unusual for short GRBs.  Both the hard and the soft components do not satisfy the Amati relation, making GRB~170817A distinctively different  from other short GRBs.
Based on gamma-ray and optical observations, we develop a model for the prompt high-energy emission associated with  GRB~170817A.  The merger of two neutron stars creates an  accretion torus of $\sim10^{-2} M_\odot$,
which supplies the black hole with magnetic flux and confines the Blandford–Znajek-powered jet.
We associate the hard prompt spike with the quasispherical breakout of the jet from the disk wind. As the jet plows through the wind with subrelativistic velocity, it creates a radiation-dominated
shock that heats the wind material to tens of kiloelectron volts, producing the soft thermal component.
\end{abstract}


\maketitle

\section{Introduction}
\label{sc:intro}

On 2017 August 17 at 12:41:04 UTC, the LIGO-Hanford detector triggered the gravitational-wave (GW) transient
GW170817~\citep{GW-trigger}. The GW signal was also found in the data of other LIGO and Virgo detectors and
was consistent with the coalescence of a binary neutron star system. Two seconds later  (UTC 2017 August 17 12:41:06) GRB~170817A was registered by GBM/\textit{Fermi} \citep{gol17,gcn21528,gcn21506} and  SPI-ACS/INTEGRAL
\citep{sav17,gcn21507}  experiments. The GBM/\textit{Fermi} localization area of GRB~170817A includes the much smaller localization region of GW170817. A search for the optical counterpart started
immediately and was carried out by a large number of ground-based facilities \citep{MMA2017}.

\section{Optical Observations}
\label{sc:optics}


The optical transient corresponding to the GW170817 and GRB~170817A was discovered independently by several
observatories. The first team to discover and report the detection of the optical counterpart was The
One-Meter, Two-Hemisphere (1M2H) group. They detected a bright uncatalogued source within the halo of the
galaxy NGC4993 in their $i$-band image obtained on 2017 August 17 23:33 UTC with the 1m Swope telescope at Las
Campanas Observatory in Chile. This source was labeled as Swope Supernova Survey 2017a \citep[SSS17a;][notice time 2017 August 17 01:05 UTC]{Swope-gcn}. SSS17a (now with the IAU designation AT2017gfo) had
coordinates RA(J2000.0) $=13^h09^m48\fs085 \pm 0\fs018$, Dec(J2000.0) $= -23\degr22\arcmin53\farcs343 \pm 0\farcs218$
and was located at a projected distance of 10\farcs6 from the center of NGC 4993, an early-type galaxy at a
distance of $\simeq40$ Mpc. Hereafter, we refer to the optical counterpart as OT.

The Distance Less Than 40~Mpc survey (DLT40; Tartaglia et al. 2017, in preparation) obtained their first image of
NGC 4993 region on 2017 August 17 23:50 UTC and independently detected the transient, automatically labeled
DLT17ck \citep[][ notice time August 18, 01:41 UTC]{DLT-gcn}.

MASTER-OAFA robotic telescope \citep{master} observed the region including NGC 4993 on August 17 23:59 UTC, and
the automated software independently detected the transient labeled MASTER OT J130948.10-232253.3
\citep[][notice time August 18, 05:38 UTC]{master-gcn}.

Visible and Infrared Survey Telescope for Astronomy (VISTA) also observed the transient SSS17a/DLT17ck/MASTER
OT J130948.10-232253.3 in the infrared band on August 18, 00:10 UT \citep[][notice time August 18, 05:04 UTC]{vista-gcn}.

Las Cumbres Observatory \citep[LCO; ][]{lco-about} surveys started observations of the \textit{Fermi} localization region
immediately after the corresponding GCN circular distribution. Approximately 5 hr later, when the LIGO-Virgo
localization map was issued, the observations were switched to the priority list of galaxies
\citep{galaxies-list}. On August 18 00:15 UTC, a new transient near NGC 4993 was detected at the position
corresponding to the OT \citep[][notice time August 18, 04:07
UTC]{lco-gcn}.

The team of DECam on the 4 m telescope of Cerro Tololo Inter-American Observatory independently
detected in optics the new source north and east of NGC 4993 in the frame taken on August 18, 00:42 UTC
\citep[][notice time August 18, 01:15 UTC]{decam-gcn}.

We started observations of the error box of GW170817 (LIGO/Virgo trigger G298048) on August 17 23:17:16 UTC
{ \citep[][notice time August 18, 14:24 UTC]{chilescope-mozaic} {with the facilities of the Chilescope
observatory (see Appendices \ref{sc:instrum} and \ref{tab:RC1000})}.
Simultaneously, we started mosaic observations of GBM/\textit{Fermi} localization area of
GRB~170817A~\citep{fermi-trigger} taking a series of spatially tiled images
with another instrument of the same observatory (see Appendix \ref{sec:ASA500}).
The observations ended $\sim5$ minutes before the GCN circular about the discovery of the transient was
distributed by Swope \citep{Swope-gcn}. The location of the NGC 4993 was out of
the covered fields of our first observational set.

We started to observe the optical counterpart of the GW170817 \citep{GW-trigger} labeled SSS17a \citep{Swope-gcn} on 2017 August 19 at 23:30:33 with RC-1000 telescope of Chilescope observatory
 \citep[][]{gcn21635} and continued observations on 2017 August 20, 21, and 24. Details of our optical observations and data reduction are presented in Appendices
 \ref{sec:SSS17a} and \ref{sec:photometry}. The final light curve in the R-filter incorporating our photometry result are presented in the Appendix \ref{sec:lightcurve}. We also compared the
 light curves of the afterglow of GRB~130603B with the light curve of the optical counterpart of GRB~170817A and found that afterglow luminosity of GRB~170817A is more than 130 times fainter than that of
 GRB~130603B in $J$-filter (see \ref{sec:lightcurve}). We  used  data of the Big Scanning Antenna (BSA) radio telescope survey at  110 MHz to   estimate the possible radio prompt emission.  Details of our radio observations are presented in
 Appendix \ref{sc:BSA}.}

\section{GRB~170817A Prompt Gamma-Ray Data Analysis}
\label{sc:data}

\subsection{\textit{SPI-ACS/INTEGRAL}}
{
The description of GRB~170817A INTEGRAL observations and detailed analysis of the burst based on INTEGRAL data are presented in \cite{sav17}. We performed a basic analysis of the SPI AntiCoincidence Shield (SPI-ACS) data with
a slightly different method to classify the burst and to estimate its energetics, and then compare our results with those of GBM/\textit{Fermi} in order to calibrate SPI-ACS consequently. We found the burst to be from a short (or Type I)
population with no precursor and extended emission components. Derived values of fluxes in energetic units are consistent with both our GBM/\textit{Fermi}, and \cite{sav17} results. The time delay between the burst onset and GW trigger was found to
be $\simeq$ 2 s. The background-subtracted light curve of the burst in SPI-ACS data with a time resolution of 0.2 s is shown in Figure~\ref{fig:acs-gbm} (panel e). The details of the analysis are presented in Appendix~\ref{sc:acs_app}.}

\begin{figure}
\includegraphics[width=\columnwidth]{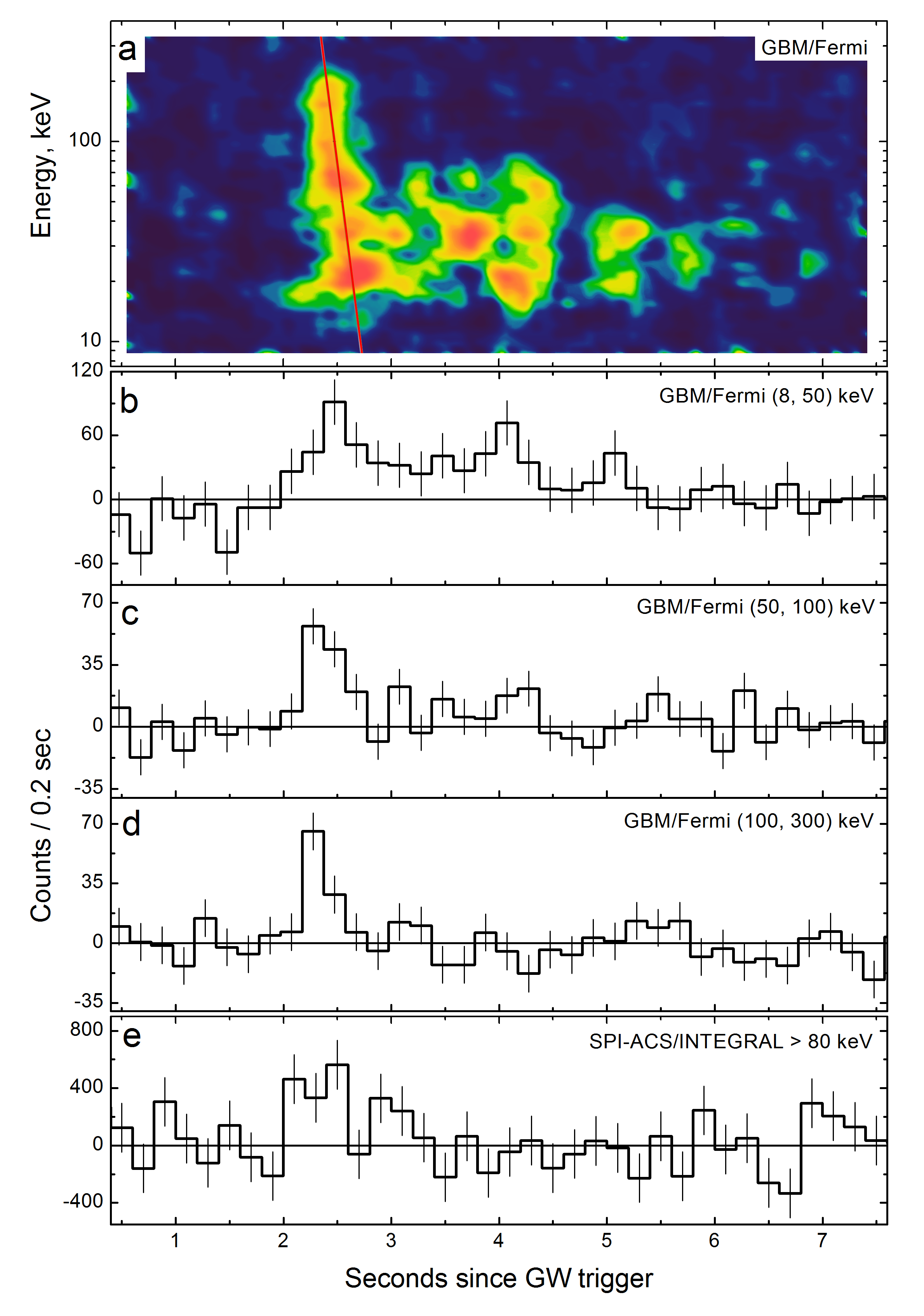}
\caption{{ Background-subtracted dynamic count spectrum (a) and light curves of GRB~170817A in data of GBM/\textit{Fermi} (NaI: 1, 2, 5) in (8, 50) keV
(b), (50, 100) keV (c), (100, 300) keV (d), and in data of SPI-ACS/INTEGRAL (e). Time in seconds since GW
trigger (UTC 2017 August 17 12:41:04) is on the X-axis, counts over 0.2 s time bins are on the Y-axis for light curves. Flux uncertainties are
presented at the 1$\sigma$ significance level. A warmer color on the dynamic count spectrum (a) corresponds to a higher flux. The two-component (pulse) structure of GRB~170817A is seen. For the first
pulse, red line represents logarithmic spectral lag behavior on energy, $lag \sim 0.24\times log(E)$. The logarithmic approximation is typical for single pulses of GRBs \citep{min14}. }}
\label{fig:acs-gbm}
\end{figure}

\subsection{GBM/\textit{Fermi}}
\label{sc:GBM}

We used the publicly accessible FTP archive \footnote{\url{ftp://legacy.gsfc.nasa.gov/fermi/data/}} as the source of the GBM/\textit{Fermi} data.

GRB~170817A was detected in GBM/\textit{Fermi} (NaI: 1, 2, 5) at the 8.7$\sigma$ significance level in the (8, 300) keV energy
range. The background-subtracted dynamic count spectrum in the (8, 400) keV range and light curves of the burst with time resolution of 0.2 s are
presented in three energy channels in panels (a)--(d) in Figure~\ref{fig:acs-gbm}. The burst onset in GBM/\textit{Fermi}
data is delayed by $\simeq2$~s compared to the GW trigger. The third-order polynomial model was
used to fit background  in time intervals (-40, -5) and (20, 70) s for all presented GBM/\textit{Fermi} light
curves.

As seen in Figure~\ref{fig:acs-gbm}, light curve of the burst consists of two different components (pulses):
the first one is short hard and the second one is visible only in soft energy range the (8, 50) keV. The two-component
structure of the burst is also confirmed by $T_{90}$ duration parameter values: in the (8, 70) keV energy range
  $T_{90}^{8-70~keV} = 2.9 \pm 0.3$ s, which is six times longer than the duration in the (70, 300) keV energy range,
$T_{90}^{70-300~keV} = 0.5 \pm 0.1$ s. \citep[That behavior cannot be explained by the well-known dependence of GRB duration on energy range, $T_{90} \propto E^{-0.4}$; see, e.g., ][]{fen95, min10b}.

\begin{table*}[]
\centering
\caption{Results of Spectral Analysis of GRB~170817A Based on GBM/\textit{Fermi} Data}
\begin{tabular}{ccccccc} \hline
Time Interval $^a$  & Model $^b$ & $\alpha$            & $E_{\rm peak}$ $^c$          & Fluence $^d$ & Hardness Ratio $^e$    & C-Stat/dof \\
 (s)                &            &                     & (keV)                    & ($10^{-7}$ erg cm$^{-2}$) &                        & \\ \hline
(-0.3, 0.3)        & PL         & $-1.50\pm0.08$         & -                       & $2.8\pm0.4$ & $0.72\pm0.13$                & 425/367 \\
 first              & BB         & -                    &  $31.6\pm3.2$              & $1.4\pm0.2$ & $2.8\pm0.6$                 & 437/367 \\
 pulse            & CPL $\star$  & $-0.9\pm0.4$ &      $230_{-80}^{+310}$    & $2.2\pm0.5$ & $1.0\pm0.2$                 & 416/366 \\ \hline
(0.8, 2.0)       & PL           & $-2.0\pm0.3$         & -                     & $1.0\pm0.4$& $0.37\pm0.15$                 & 439/367 \\
 second           & BB $\star$   & -                    &  $11.2\pm1.5$         & $0.68\pm0.11$ & $0.24\pm0.09$               & 422/367 \\
  pulse           & CPL          & $2.3_{-1.9}^{+4.5}$  & $43_{-7}^{+9}$  & $0.67\pm0.12$  & $0.23\pm0.10$                & 422/366 \\ \hline
(-0.3, 2.0)       & PL           & $-1.8\pm0.1$         & -                     & $3.7\pm0.7$& $0.49\pm0.09$               & 450/367\\
 whole            & BB           & -                    & $13.3\pm1.3$          & $1.7\pm0.2$ & $0.39\pm0.08$               & 447/367 \\
 burst            & CPL $\star$  & $-0.5_{-0.7}^{+0.9}$ & $65_{-14}^{+35}$    & $2.1\pm0.3$ & $0.46\pm0.09$               & 440/366 \\\hline

\multicolumn{7}{l}{$^{a}$ Time interval relative to GBM trigger (UTC 2017 August 17 12:41:06.475).}\\
\multicolumn{7}{l}{$^{b}$ PL is for power law, CPL is for power law with an exponential cutoff, and BB is for a
thermal model.}\\
\multicolumn{7}{l}{$^{c}$ For the BB spectral model, the $E_{\rm peak}$ column gives the parameter kT.}\\
\multicolumn{7}{l}{$^{d}$ Fluence in the (10, 1000) keV energy range.}\\
\multicolumn{7}{l}{$^{e}$ Hardness ratio calculated as a ratio of photon fluxes in the (50, 300) keV and the (15, 50)
keV energy bands.}\\
\multicolumn{7}{l}{$\star$ Optimal spectral model.}\\

\label{tab:gbmspectra}
\end{tabular}
\end{table*}

To construct and fit the energy spectra, we used the RMfit v4.3.2 software package\footnote{\url{http://fermi.gsfc.nasa.gov/ssc/data/analysis/rmfit/}} developed to
analyze the GBM data of the \textit{Fermi} observatory. The
method of spectral analysis
is similar to that proposed by \cite{gruber14}, who also used the RMfit software package. To
fit the energy spectra and to choose an optimal spectral model, we used modified Cash statistics \citep[C-Stat; see ][]{cash79}.

Results of spectral analysis performed for GRB~170817A using data of GBM/\textit{Fermi} (NaI: 1, 2, 5, BGO: 0) are summarized
in Table~\ref{tab:gbmspectra}. We analyzed three time intervals covering the whole burst --- (-0.3, 2.0) s since
GBM trigger, the first hard pulse (-0.3, 0.3) s, and the second soft pulse (0.8, 2.0) s --- with three spectral
models --- power law (PL), power law with exponential cutoff (CPL), and thermal model (BB). {
To choose the optimal spectral model, we used the $\Delta$C-Stat $ > \Delta$C-Stat$_{\rm crit}$ criterion, where $\Delta$C-Stat is
the difference between C-Stat values obtained for various models (the last column in Table~\ref{tab:gbmspectra}) and
$\Delta$C-Stat$_{\rm crit} \simeq 8.5$ was obtained via simulations in \cite{gruber14} for PL versus CPL model comparison.}

The energetic spectrum of the whole burst (time interval (-0.3, 2.0) s) is best fitted by the CPL model with $\alpha$
= $-0.5_{-0.7}^{+0.9}$ and $E_{\rm peak}$ = $65_{-14}^{+35}$ keV (the improvement over PL model is $\Delta$C-Stat = 10). The fluence in the (10, 1000) keV range is $F = (2.1\pm0.3)\times10^{-7}$ erg cm$^{-2}$. Using the
CPL spectral model, we calculated the hardness ratio of photon fluxes between (50, 300) keV and (15, 50) keV energy
bands, HR = $0.46\pm0.09$, which, along with duration $T_{90}^{70-300~keV} = 0.5
\pm 0.1$~s, characterizes the burst to be a typical short but soft one \citep[e.g., see Fig. 6 in ][]{kienlin14}.
The probability that GRB~170817A is from the short population was estimated using duration and hardness values in
\cite{gol17} and found to be $P \simeq 73\%$.

The optimal spectral model for the first pulse (time interval (-0.3, 0.3) s) of the GRB~170817A is the CPL model
(Table~\ref{tab:gbmspectra}) with $\alpha$ = $-0.9\pm0.4$ and $E_{\rm peak}$ = $230_{-80}^{+310}$ keV (the improvement over the PL model is $\Delta$C-Stat = 9). Using the CPL spectral model, we calculated a
hardness ratio between (50, 300) keV and (15, 50) keV energy bands, HR = $1.0\pm0.2$, which describes the
first pulse as a hard one \citep[see Fig. 6 in ][]{kienlin14}.

The optimal spectral model for the second pulse (time interval (0.8, 3.0) s) of the GRB~170817A is a thermal
model with $kT = 11.2\pm1.5$ keV and the fluence is in the (10, 1000) keV energy range of $F = (6.8\pm1.1)\times10^{-8}$
erg cm$^{-2}$ (Table~\ref{tab:gbmspectra}). The CPL model gives the same goodness of fit, but with a positive value
of $\alpha$ = $2.3_{-1.9}^{+4.5}$, bringing this model maximally close to Wien's law $N_E \sim E^2 exp(-E/kT)$ --- the spectral shape that the radiation
originating in an optically thick hot medium whose opacity is dominated by Compton scattering off electrons acquires \citep{komp56},
which also confirms the thermal nature of the second pulse.  The $E_{\rm peak}$ parameter was found to be $E_{\rm peak} = 43_{-7}^{+9}$
keV within the CPL model. The improvement of C-Stat for CPL and BB models comparing to the PL one is $\Delta$C-Stat = 17, which is two times larger
than $\Delta$C-Stat$_{\rm crit} \simeq 8.5$ found in \cite{gruber14} and used as model selection criterion. Within the CPL model, the hardness ratio between (50, 300) keV and (15, 50) keV energy bands,
HR = $0.23\pm0.10$, which describes the second pulse to be a very soft one (see Fig. 6 in \cite{kienlin14}).

\begin{figure}
\includegraphics[width=\columnwidth]{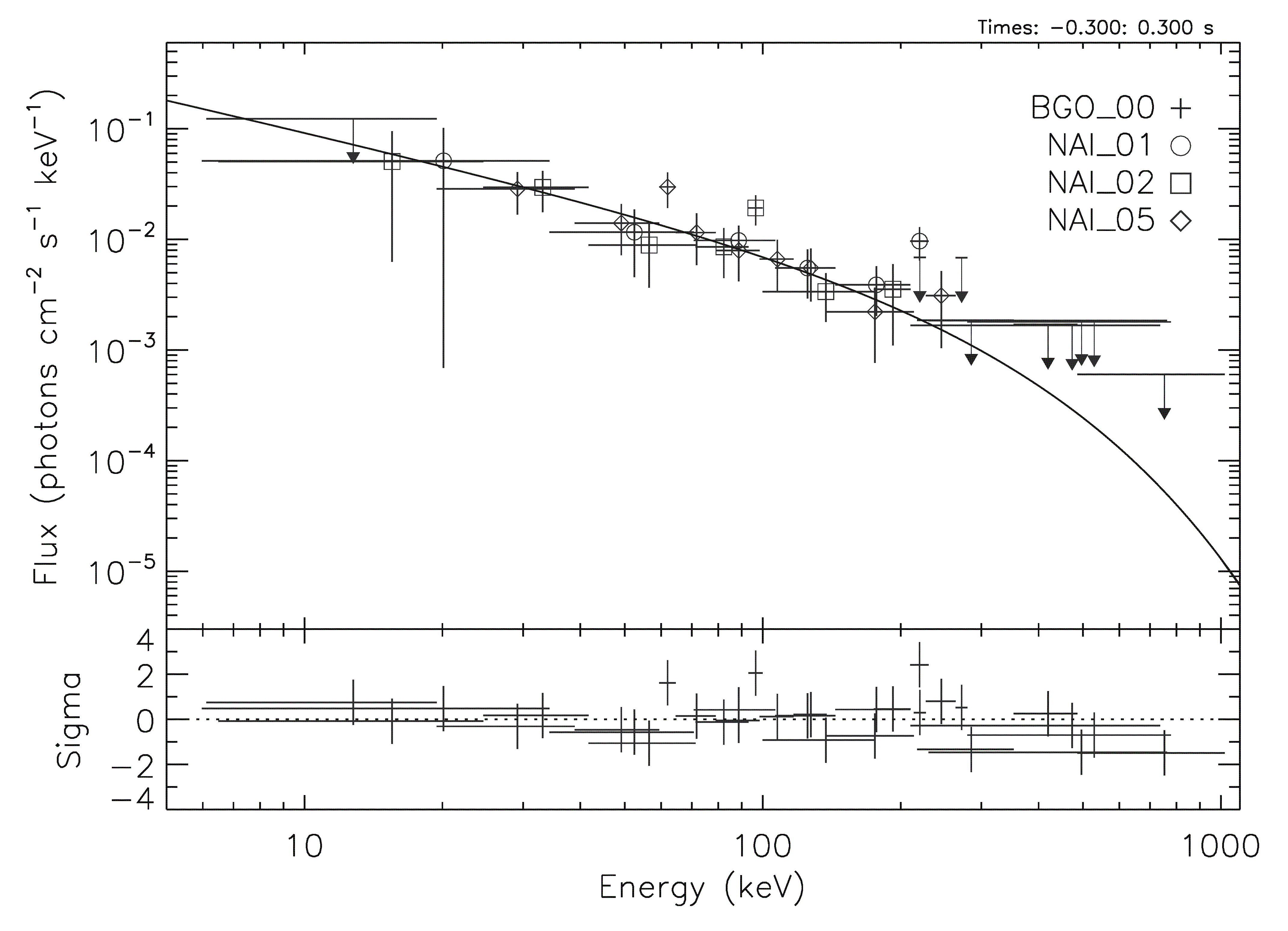}
\caption{Photon spectrum (upper panel) and residuals (lower panel) of the CPL spectrum model for the first pulse.}
\label{fig:1CPL}
\end{figure}

\begin{figure}
\includegraphics[width=\columnwidth]{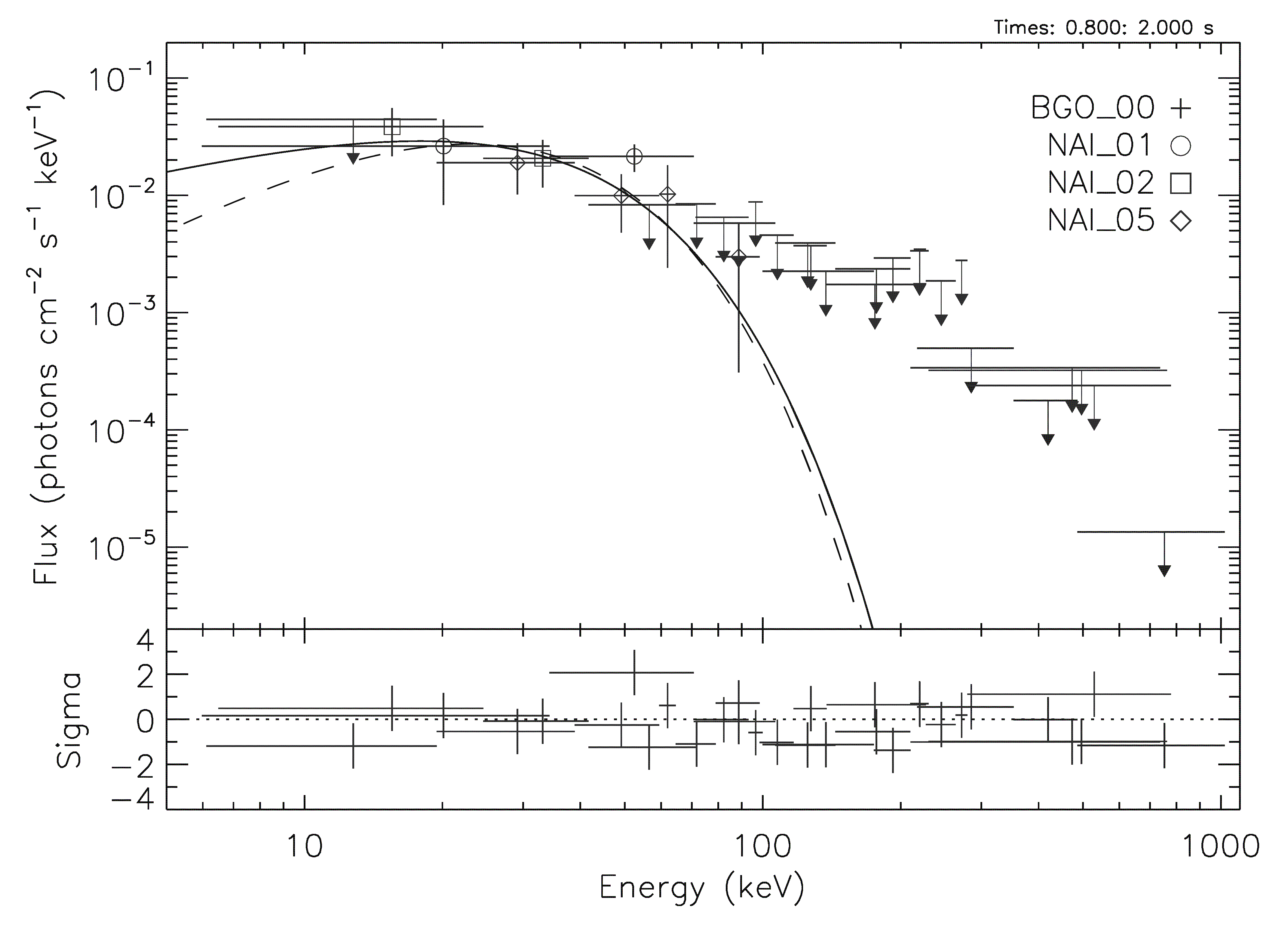}
\caption{Photon spectrum (upper panel) and residuals (lower panel) of the black body spectrum model for the second pulse. The dashed line represents the CPL fit of the photon spectrum.}
\label{fig:2BB}
\end{figure}

\begin{figure}
\includegraphics[width=\columnwidth]{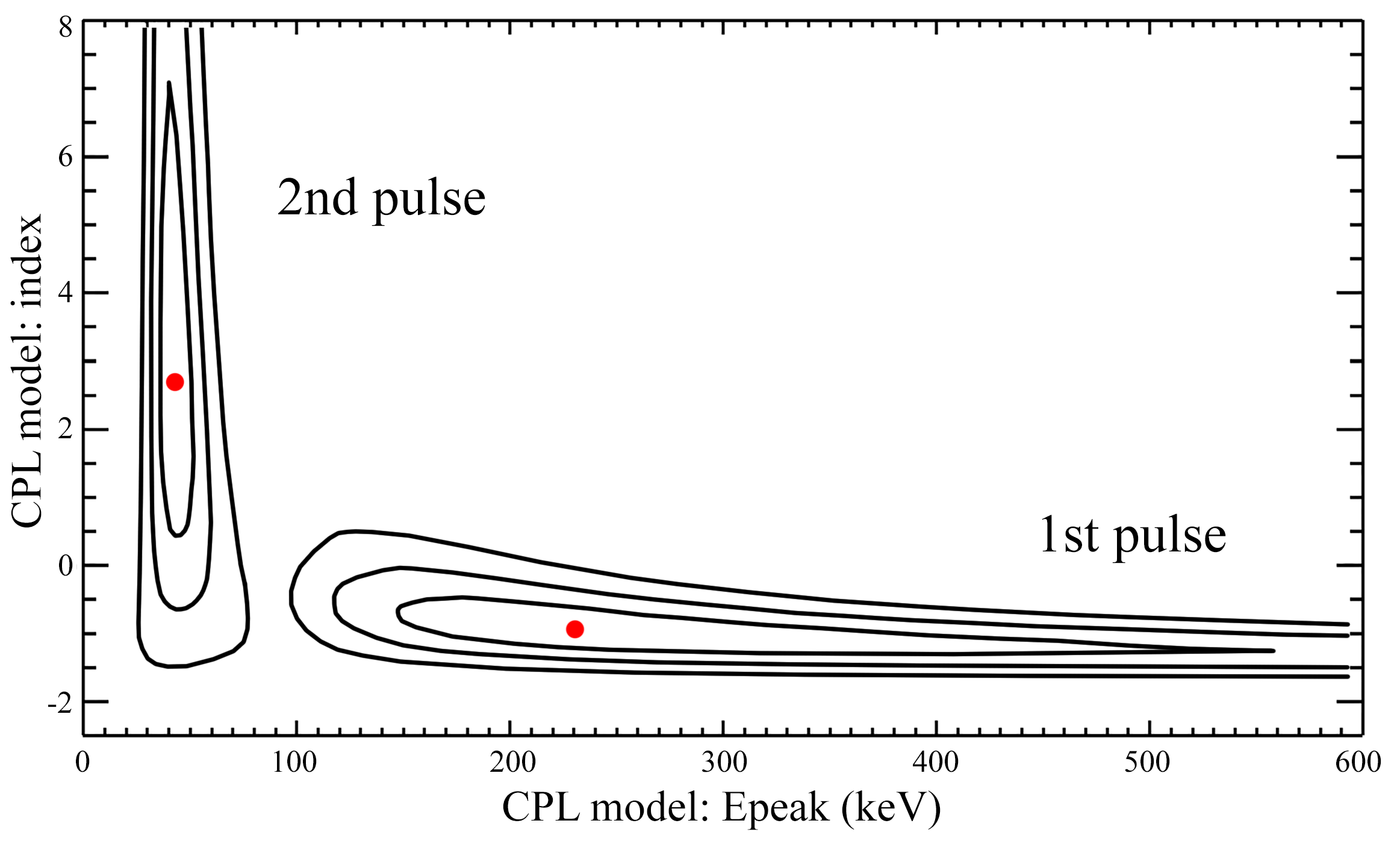}
\caption{Confidence regions at the 1, 2, 3$\sigma$ levels of the CPL spectrum parameters (power-law index, $\alpha$ and $E_{\rm peak}$) for the first and second pulses. }
\label{fig:confidence}
\end{figure}

Optimal spectral models are presented in Figures~\ref{fig:1CPL} and \ref{fig:2BB}, while confidence regions of the parameters of CPL models for both pulses are presented in Figure~\ref{fig:confidence}. As can be seen from
Figure~\ref{fig:confidence}, the spectrum describing these two pulses is not evolving, but rather is a manifestation of the different nature of the emission. The results of our spectral analysis are comparable with ones presented in \cite{gol17}.

We did not find any precursor or extended emission components in GBM/\textit{Fermi} data at time scales from 0.1 up to
20 s in the time interval (-10, 30) s since GBM triggers in the energy range of (8, 300) keV. To estimate upper
limits on their intensity in energetic units, we derived a conversion factor in the (8, 300) keV range using count fluence of the burst and
energetic fluence within the CPL spectral model (Table~\ref{tab:gbmspectra}). We found that 1 count in GBM/\textit{Fermi}
corresponds to $\sim 3.2\times10^{-10} $ erg cm$^{-2}$ in an energy range of (8, 300) keV. At a time scale of 0.1~s, the upper
limit on precursor activity is $S_{\rm prec} \simeq 1.6\times10^{-8} $ erg cm$^{-2}$ at the 3$\sigma$ significance level.
At a time scale of 20 s, upper limit on extended emission activity is $S_{\rm EE} \simeq 2.3\times10^{-7} $
erg cm$^{-2}$ at a the 3$\sigma$ significance level. The upper limits are two times deeper than ones obtained with SPI-ACS (see Appendix~\ref{sc:acs_app}).

A spectral lag analysis based on the CCF method \citep[see, e.g., ][]{min14} was performed for GRB~170817A using the light curves
detected by GBM/\textit{Fermi} (NaI: 1, 2, 5) in the energy bands (8, 50) keV, (50, 100) keV,
and (100, 300) keV with a time resolution of 0.04 s. The spectral lag between (8, 50) keV and (50, 100) keV bands is
negligible, lag = $0.01\pm0.08$ s, but between (8, 50) keV and (100, 300) keV lag is significant, lag =
$0.27\pm0.05$ s. It might represent the hard-to-soft spectral evolution of the first pulse (see
Figure~\ref{fig:acs-gbm}, upper panel).

\subsection{Amati Relation}

\begin{table*}[]
\centering
\caption{GRB~170817A energetics}
\begin{tabular}{cccccc} \hline
Time Interval $^a$  & Spectral   &  Flux $^c$                  &  Fluence $^d$           & $E_{\rm iso}$      & $L_{\rm iso}$        \\
 (s)                & model~$^b$  & $10^{-7}$~erg/(cm$^{2}$~s)  &  $10^{-7}$~erg~cm$^{-2}$ & $10^{46}$~erg  &  $10^{46}$~erg~s$^{-1}$  \\ \hline
(-0.3, 0.3)        & CPL         & $3.8\pm0.9$               & $2.3\pm0.9$             & $4.9\pm1.2$    & $8.1\pm1.9$       \\
(0.8, 2.0)         & BB          & $0.59\pm0.10$               & $0.71\pm0.12$         & $1.5\pm0.3$    & $1.3\pm0.2$      \\
(0.8, 2.0)         & CPL         & $0.56\pm0.10$               & $0.67\pm0.13$         & $1.4\pm0.3$    & $1.2\pm0.2$      \\
(-0.3, 2.0)         & CPL         & $0.96\pm0.14$               & $2.2\pm0.32$            & $4.7\pm0.7$    & $2.1\pm0.3$       \\\hline

\multicolumn{6}{l}{$^{a}$ Relative to the GBM trigger (UTC 2017 August 17 12:41:06.475).}\\
\multicolumn{6}{l}{$^{b}$ CPL is for power law with an exponential cutoff, BB for a thermal model.}\\
\multicolumn{6}{l}{$^{c}$ Energy flux in the (1, 10,000) keV energy range.}\\
\multicolumn{6}{l}{$^{d}$ Energy fluence in the (1, 10,000) keV energy range.}\\

\label{tab:gbmenergy}
\end{tabular}
\end{table*}

One of the interesting phenomenological relations for GRBs is the Amati diagram, i.e., the dependence of the
equivalent isotropic energy $E_{\rm iso}$ emitted in gamma-rays between (1, 10,000) keV on parameter $E_{\rm peak}(1 +
z)$ in the source frame \citep{amati_10}, which can also be used for the classification of GRBs \citep[see, e.g., ][]{qin13}.

\begin{figure}
\includegraphics[width=\columnwidth]{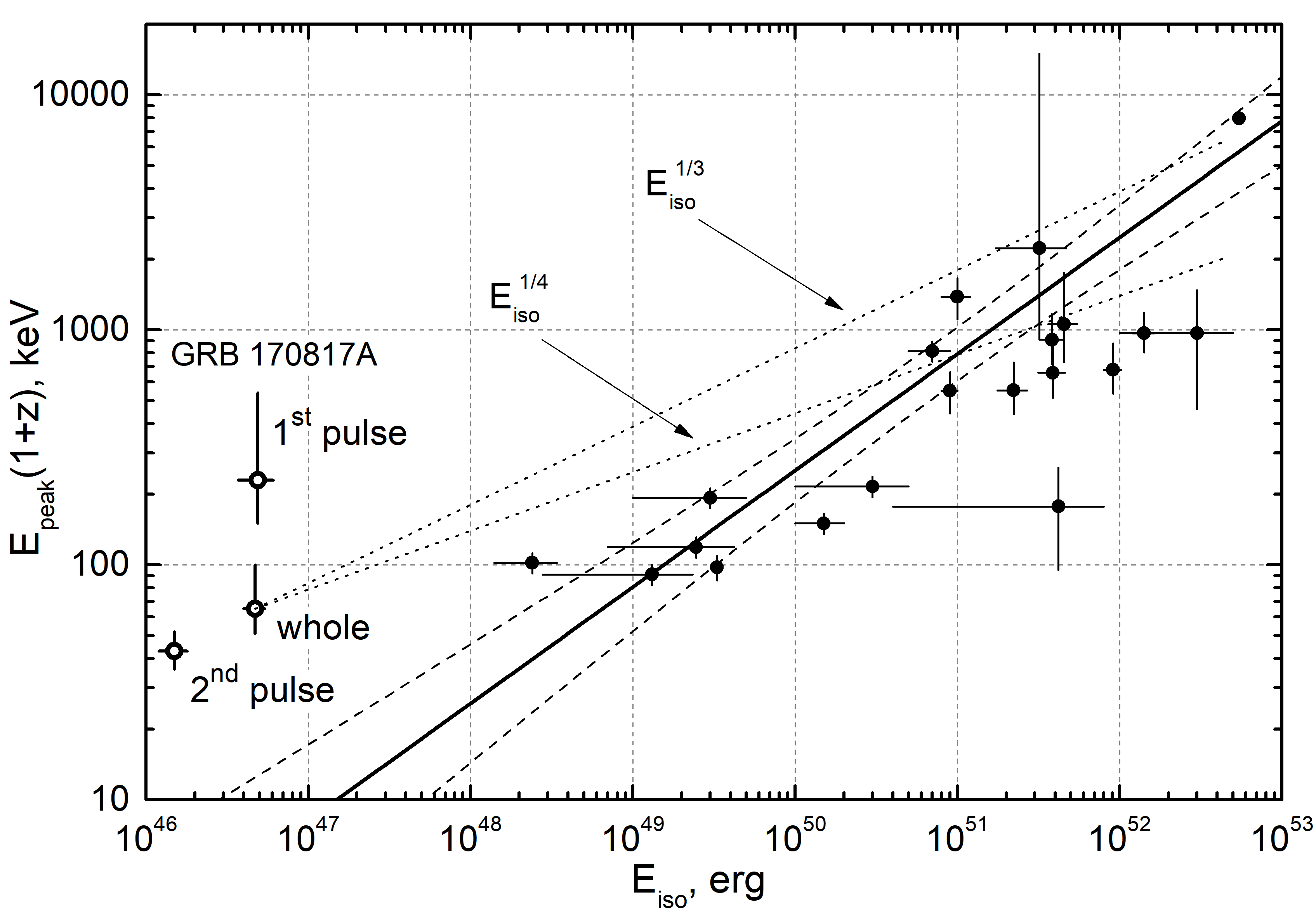}
\caption{Amati diagram; the dependence of the equivalent isotropic energy $E_{\rm iso}$ emitted in the (1, 10000) keV range on $E_{\rm peak}(1 + z)$ in the source frame \citep{amati_10}, for short bursts. A power-law fit (thick solid line) with dashed lines bounding the 2$\sigma$ region are shown. Dotted lines starting from the point for the whole GRB~170817A indicate the dependence $E_{\rm peak}(1 + z) \sim E_{\rm iso}^{1/3}$ and $E_{\rm peak}(1 + z) \sim E_{\rm iso}^{1/4}$. Uncertainties for $E_{\rm iso}$ and $E_{\rm peak}(1 + z)$ are presented at the 1$\sigma$ significance level.}
\label{fig:amati}
\end{figure}

Using the experimental data given in \citet{svi16,qin13,min17} we constructed the Amati diagram for 20 short bursts and fitted the relation with a power-law-like logarithmic model: {
$\log\left(\frac{E_{\rm peak}(1 + z)}{1 \;\rm keV}\right) = (0.50 \pm 0.04) \log \left(\frac{E_{\rm iso}}{1 \rm erg}\right) + (-22.4 \pm 1.8)$. Details on the investigation of the Amati relation for short GRBs can be found elsewhere (P. Minaev 2018, in preparation).

Using optimal spectral models derived in the previous subsection for GRB~170817A  (Table~\ref{tab:gbmspectra}) and assuming a redshift of z = 0.00968 and a luminosity distance of $D_{L}$ = 42.0 Mpc for the source  \citep{2009MNRAS.399..683J},
we estimated $E_{\rm iso}$ and $L_{\rm iso}$ parameters in the (1, 10,000) keV range
(Table~\ref{tab:gbmenergy}). All burst components (first, second pulses, and the whole burst) lie far above the 2$\sigma$ correlation region (Figure~\ref{fig:amati}).

If we assume that the burst GRB~170817A should obey the Amati relation, we can draw a trajectory of the burst parameters as $E_{p} \sim E_{\rm iso}^{1/3}$ if it is a cone relativistic jet emission, and  $E_{p} \sim E_{\rm iso}^{1/4}$ if it is a spherical relativistic emission \citep[see, e.g., ][]{eic04,lev05}. One can see that neither of the trajectories cross the Amati relation at a reasonable $E_{\rm iso}$, especially for the first pulse. This strongly supports an alternative explanation for the nature of the first pulse as well as the whole burst in comparison with the usual short GRBs presented in the Amati relation (Figure~\ref{fig:amati}).
}

\section{The Scenario}
\label{sc:model}

\begin{figure}
 \begin{center}
\includegraphics[width=.99\linewidth]{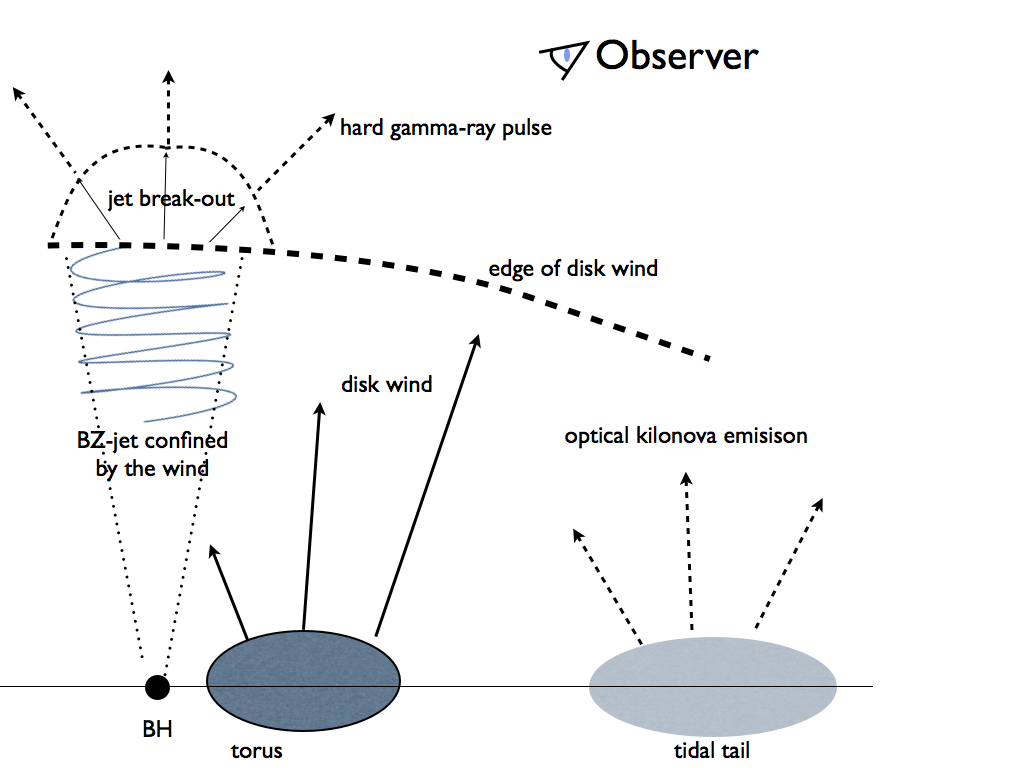}
\end{center}
\caption{Sketch of our scenario. After merging of NSs, a BH is formed, which is surrounded by a compact accretion disk with {\bf intense} wind.
After  sufficient magnetic flux is  accumulated on the BH an electromagnetic jet is launched, confined initially by the disk wind. After propagating with mildly relativistic velocities,  the jet  breaks out from the wind zone in a semi-isotropic fashion, reaching highly relativistic {\Lf}s. This is the prompt GRB. }
\label{fig:SK}
\end{figure}

\begin{figure}
\includegraphics[width=\columnwidth]{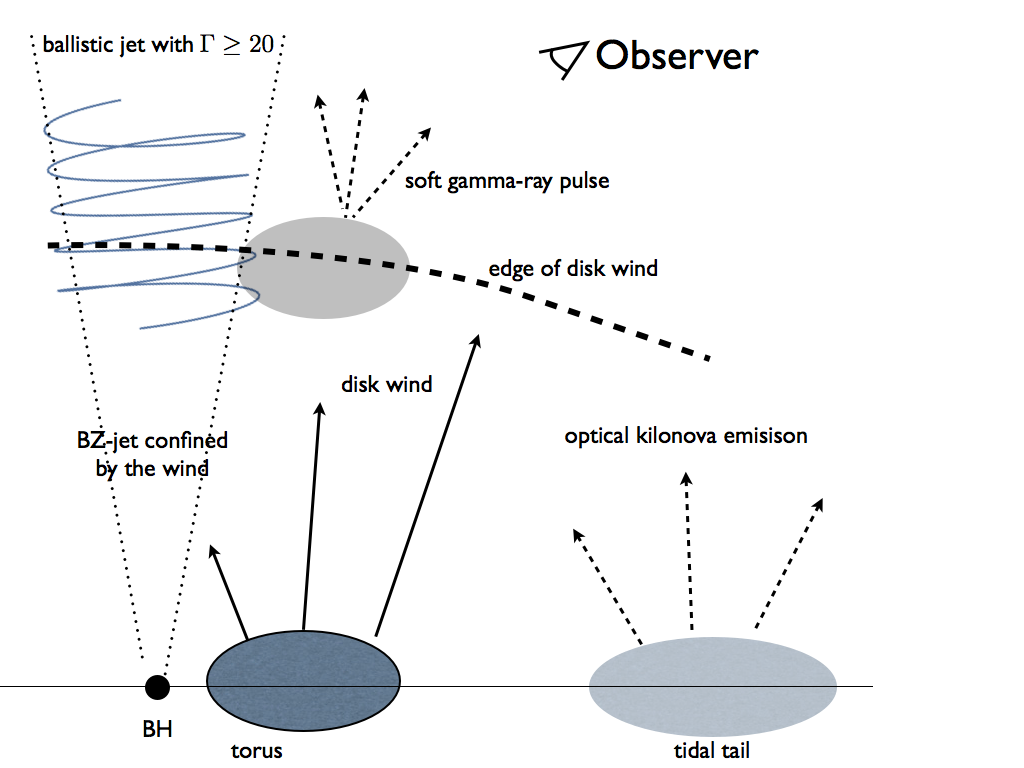}
\caption{Post-breakout structure of the flow. The wind from the disk keeps collimating the jet for a few seconds. The jet propagates ballistically with  {\Lf}s $\geq 20$ and is unobservable due to deboosting. The shocked wind material produces  thermal extended gamma-ray component with $T \sim 10$ keV.}
\label{picture-GRB-GW003}
\end{figure}

The key points of the model are  illustrated in Figures \ref{fig:SK}-\ref{picture-GRB-GW003}.

The detection of the EM signal contemporaneous with gravitational waves is  consistent with the binary NS scenario for short GRBs \citep{bnpp84,p86,1989Natur.340..126E}.
Qualitatively, the evolution of merging \NSs\ follows a well-defined path \citep[\eg][and many others]{rgl16,rlps16,bairez17}, though many details, like   the effects of different equations of states,
various mass ratios, initial spins, and \Bf\ evolution,  remain to be settled. An active stage of a merger lasts $\sim 10-100$ milliseconds after which the {\NS}s collapse into BH.
\footnote{ We disfavor alternative  scenarios, like the formation of a supermassive fast rotating  neutron star with a strong magnetic field \citep{bk70,LW70,U92,MBA06,2008MNRAS.385.1455M}, so we assume the collapse of the very massive \NS{} to a BH.}
The BH is fairly fast rotating with the Kerr parameter $a \sim 0.7$ \citep{rgl16,rlps16}. The mass of the resulting BH is  somewhat smaller than the sum of masses due to emission of neutrinos,
gravitational waves, ejection of the tidal disruption tail, and wind from the accretion disk, so we can estimate it as $M_{\rm BH}\approx 2.5 M_\odot$. The amount of the ejected material is especially uncertain, but is of utmost importance for the production of the EM signal.
It is expected that, first, tidal tails eject  $0.01-0.03 M_{\odot}$ with outflow speed $v_{\rm ex}\approx 0.1-0.3 c$. This material is likely a site of r-process nucleosynthesis and
can be seen as a kilonova --- an optical emission of peak luminosity $\sim 10^{41}$ erg s $^{-1}$ lasting for a few weeks \citep[see ][]{rkl11,bk13,bgj13,kbb13,gkr14,wsn14,bkw16,wtjj17}.

Second, during the merger,  an accretion torus of  $\sim 0.01 M_{\odot}$ forms around the BH with viscous time  0.1~s \citep{rgl16,rlps16,bairez17}.
Highly different masses of merging NSs are good for the formation of a massive disk, which can enhance jet power and aid its confinement.
This accretion disk plays the most important part in our model. For a few seconds, the disk undergoes viscous spreading, produces powerful
neutrino-driven winds, and exercises accretion onto the BH via  the inner edge (presumably close to the innermost stable orbit), see Section \ref{disk}.
Interestingly,  in the case of BH+NS merging, the neutrino heating mechanism \citep{elps89,bir07,zb11} can be the main source of the jet's energy \citep[see][]{bp11}.
In the case of NS+NS merging, the mass of the accretion torus is relatively small ($<10^{-2}M_{\odot}$), so the neutrino heating mechanism cannot be efficient on time scales of a few seconds.

At the same time, \Bfs\ are amplified within the disk to $\sim 10^{15}$ G \citep{2011ApJ...732L...6R} due to  the development of  MRI and the presence of the velocity shear.
As the matter is accreted onto the BH, the BH accumulates magnetic flux.  Accumulation of the magnetic flux leads to a delay for the jet to switch-on. At the same time, baryons
slide off into the BH along \Bf\ lines, leaving  polar  regions with low density. This creates conditions favorable for the operation of the  Blandford--Znajek (BZ) mechanism \citep{bz77,KB09}.

The BZ mechanism -- production of EM outflows from the \Bf\ supplied by the disk -- also requires the presence of the external medium to produce jets (collimated outflows).
It is the heavy baryonic wind from the disk that provides the required collimating surrounding. Importantly, the disk wind has only limited spatial extent --
propagating with a velocity of $\sim 0.1 c$ for about a few seconds it reaches out only to a few $\times 10^9$ cm. Outside of the wind, the surrounding is very clean.

The  BZ jet propagates through a dense wind with mildly relativistic velocity (Section \ref{propagate}). At the same time, the jet is mildly dissipative (Section \ref{spike}; and thus highly optically thick).
In a few seconds, the head of the EM jet reaches the edge of the wind. During the breakout, the head part of the jet expands with highly relativistic velocities in the direction of jet propagation and
moderately relativistic velocities to the sides.
Sudden expansion leads to pair annihilation and the production of gamma-ray emission in  a manner similar to conventional
models of GRBs (Section \ref{spike}). This is the prompt GRB.

After the jet breaks out from the wind, a rarefaction wave propagating toward the BH leads to the acceleration of the jet, which loses internal causal contact and
propagates nearly ballistically (Figure \ref{picture-GRB-GW003}). Since the jet emission is highly debeamed, the jet becomes unobservable. The wind emission shocked
by the breaking-out jet produces the soft tail (Figure \ref{picture-GRB-GW003}).
A similar model for the formation of the first hard gamma-ray pulse was independently suggested in the paper by \cite{gnph17}. This paper  focuses on the effects of the jet breakout, both hydrodynamic and radiation processes.

\subsection{Jet Breakout -- Prompt Hard GRB}
\label{spike}

The BZ mechanism of jet launching requires clean plasma -- there should be little mass loading and magnetic dissipation close to the source.
This is indeed expected if the flow originates on field lines that penetrate the BH. It is expected that a jet launched by the BH is
highly magnetized, $\sigma \gg 1$. Importantly,   the confined jet with $\sigma \gg 1$ that propagates with non-relativistic velocity
must necessarily be dissipative  \citep{2006NJPh....8..119L}.  Briefly, the magnetic flux and energy are supplied to the
inflating bubble by a rate that cannot be accommodated within the bubble \citep[this problem is also known as the $\sigma$-problem in pulsar winds; ][]{1974MNRAS.167....1R,kc84}.
If dissipation becomes important, it
will destroy a significant fraction of the magnetic energy and most importantly the toroidal flux will be eliminated.
Latest 3D numerical simulations of pulsar wind nebulae indeed demonstrate that magnetic flux (and some magnetic energy) are
dissipated within non-relativistically expanding cavities, thus resolving the $\sigma$-problem \citep{2013MNRAS.431L..48P,2014MNRAS.438..278P}.
A considerable fraction of the magnetic energy that has been produced before the jet breakout is dissipated.

At the time of the jet breakout its magnetization is not very high. At the breakout,
the highly over-pressured jet loses both radial and lateral confinement. At this point, the  jet dynamics changes in two important ways.
First, the leading part of the jet expands nearly spherically, within  a solid angle $\sim 2 \pi$,  with large {\Lf}s
\citep[see][for discussions of breakout dynamics in relativistic magnetized outflows; also,
in Appendix \ref{ap:bomb}, we construct a simple model  of a breakout of a magnetically dominated jet -- the  ``force-free magnetic bomb'']{2010PhRvE..82e6305L,2012PhRvE..85b6401L}; the dynamics of  non-magnetized relativistic breakout flows was considered by
\cite{JohnsonMcKee,2012ApJ...747...88N}; for discussions of breakout dynamics in core collapse supernovae, see,  \eg\,
\cite{1976ApJS...32..233W,2001ApJ...551..946T,1992ApJ...393..742E}.  At this point, the dynamics of pair-loaded magnetized outflows will resemble
the conventional  picture of magnetized GRB outflows \citep[see, \eg\, early discussion by ][]{LyutikovUsov,2017ApJ...838..125B}.

We suggest that this nearly spherical, highly relativistic outflow  produces the prompt GRB spike. As the optically thick jet breaks out from the confining disk wind,
the pair density falls out of equilibrium. Comptonization with the mildly optically thick region produces the observed GRB emission,
similar to the photospheric emission in conventional GRB outflow \citep[][]{2010MNRAS.407.1033B,2017SSRv..207...87B}. The emission
is expected to peak at energies $\sim \Gamma \epsilon_{\rm ph}\geq 400 $  keV, where $\epsilon_{\rm ph} \sim 20 $ keV is the photospheric
temperature in relativistic outflows \citep{Paczynski86} and $\Gamma \geq 25$ is the bulk \Lf, see Eq.  (\ref{lc}).
Relativistic hydrodynamical simulations and calculation of jet brightness from a viewing angle can be found in \cite{llc17}.

Second, the bulk of the  jet  accelerates along the jet axis -- there is no more wind material to plow through. As the bulk of the  jet accelerates to
$\Gamma \gg 1$ (see (\ref{lc})), its propagation becomes nearly ballistic.

\section{Outflow Parameters Inferred from Observational Constraints}

\subsection{Lorentz Factor for the Prompt Spike}
For the hard prompt emission, using the total fluence of $2.3 \times 10^{-7} \rm erg \mbox{ cm}^{-2}$, the total energetics estimates to $\sim 4.9 \times 10^{46}$ erg, with peak power
$\sim 10^{47}$ erg s$^{-1}$. The corresponding compactness parameter is large
\be
l_c =  \frac{L _\gamma \sigma_T \xi}{m_e c ^4 t} = 3 \times 10^8 \xi t_{-0.3}^{-1},
\label{lc}
\ee
here $\xi$ is the fraction of photons above the pair production threshold.
Since   the optical depth  to pair production is $\propto l_c/\Gamma^6$ \citep[\eg][]{goodman86,2001ApJ...555..540L},  it is required  that the flow producing the prompt burst accelerates
to
\be
\Gamma_{\rm j, min} = \left(\frac{\sigma_{\rm T} L_{\rm ht}\xi}{m_{\rm e}c^4 t}\right)^{1/6} \approx 25\xi^{1/6},
\label{eq:gjmin}
\ee
unfortunately, we have no information about the number of high-energy photons and our estimation Equation~(\ref{eq:gjmin}) can be significantly reduced.
This in turn requires the emitting  jet to be sufficiently clean (lacks baryons). This is an important constraint since the neutron-rich
environment is expected to pollute the jet \citep{1999ApJ...521..640D,2003ApJ...588..931B}.

\subsection{Opening Angle of the Jet from the Soft Gamma-Ray Component}
\label{Lee}

Properties of the soft  thermal emission can be used to infer the geometrical properties of the outflow.
In our model, the soft gamma-ray component originates from the wind material shocked by the jet and thus should
have a physical size of the order of the jet  radius at the moment of breakout.
The total (thermal) luminosity of the second soft component is $L_{\rm ee} = 1.3 \times 10^{46}$~erg s$^{-1}$. Thus,
the
emitting area is
\be
S= \frac{L_{\rm ee}}{\sigma_{SB} T_{\rm ee}^4} = 1.1\times 10^{18}\mbox{cm}^2
\label{eq:sjh}
\ee
and the radius of emitting region $r_{\rm S}$ is
\be
r_{\rm S}= \sqrt{\frac{S}{\pi}} =6 \times 10^8 \mbox{cm}.
\ee
If the wind from the torus propagates  with velocity $v_{\rm w} \approx0.1c$ \citep[see ][]{bb11}, then in 2.4 s it would expand to $r_{\rm w} \approx 7 \times 10^{9}$ cm.
Thus, the jet opening angle is
\be
\theta_{\rm j} = \frac{r_{\rm S}}{r_{\rm w}} \approx 0.08.
\label{eq:thetaj}
\ee
This value agrees well with results of GRMHD simulations
\citep[see ][]{b08,bb11,kbg17}.

\subsection{Magnetic Field at the Source}

As the jet breaks out from the wind, the quasispherical expansion of the jet material will form a fireball that resembles the pair-loaded fireball discussed  earlier in
the literature \citep{pacz90,rm05}. A fraction $1/\eta_{\rm em}\ll 1 $ of the energy stored in the wind prior to the jet-break will be radiated ($\eta_{\rm em}$  depends on parameters of outflow like the energy injection rate, magnetization, and
baryonic loading \citep[\eg][]{LyutikovUsov}.

We can relate the  observed flux in the first/hard peak  to  the jet head energy density as
 \be
 u_{\rm em}= \eta_{\rm em} \frac{ 4 L_{\rm peak}} {S c}.
 \label{uem}
 \ee
 This can be used to estimate the equipartition \Bf\  $B_{\rm eq}$ in the jet:
 \be
 B_{\rm eq}= \frac{\sqrt{8 \pi u_{\rm em}} }{\sqrt{3}} \approx 3\times 10^{10} \eta_{\rm em,1}^{1/2} \mbox{ G}.
 \ee
  If originating  from a BH with the surface field $B_{\rm BH}$, the \Bf\  at the emission site is
 \be
  B_{\rm eq} = B_{BH} \left( \frac{r_{\rm g}}{R_{\rm LC}}\right)^2 \frac{R_{\rm LC}}{r_{\rm S}},
  \ee
where $r_{\rm g}$ is BH Kerr radius, $R_{\rm C} \sim 4 r_{\rm g}/ a_{\rm BH}$ is the \LC\ radius, and $a_{\rm BH}$ is the Kerr parameter. (Scaling is different from the NS case, here it is $R_{\rm LC}^2$ not $R_{\rm LC}^4$.)

  Thus, the \Bf\ on the surface of the BH can be estimated as
  \be
   B_{\rm BH} \sim \frac{4}{a_{\rm BH}}\frac{r_{\rm S}}{r_{\rm g}} B_{\rm eq}  \sim  2\times 10^{14} \eta_{\rm em,1}^{1/2} a_{\rm BH,-0.2}^{-1} \mbox{ G},
   \label{BBH}
   \ee
   where we scaled the Kerr parameter to $0.7$.
      These fields are  higher, by about one or two orders of magnitude than is expected from  surface \Bfs\ of  merging  \NSs, indicating that a dynamo process was operational during or after the merger (either on the collapsing \NS\ or in the surrounding torus).

Substituting \Bf\ from Eq.~(\ref{BBH}) to the equation of jet power
\be
L_{\rm BZ}  \approx \frac{ B_{\rm BH}^2 r_{\rm g}^2 c}{24}f_{\rm BH}(a_{\rm BH}) ,
\label{eq:LBZi}
\ee
here $0\le f_{\rm BH}(a_{\rm BH})<1$ is a factor that depends on the BH spin parameter  \citep[see more details in][]{bz77,bk08b}, we obtain expected power  as
      \be
      L_j \sim  7 \eta_{\rm em} L_{\rm peak} \sim  10^{49} \eta_{\rm em,1} \mbox{ erg s}^{-1}.
      \label{Lj}
      \ee
      To reiterate, Equation (\ref{Lj}) related the power $L_{\rm ee}$ of the hard gamma-ray emission to the power of the BZ jet $L_j$.
      The estimate  (\ref{Lj}) compares favorably to the estimate of the jet power derived from modeling the evolution of the accretion torus,  Equation (\ref{eq:LBZ}), conversion parameter
      $\eta_{\rm em} \approx 10$, which corresponds to a jet power on of $10^{49}\mbox{ erg s}^{-1}$ \citep[see][]{pacz90}.

\section{The Model}

\subsection{Evolution of the Accretion Disk}
\label{disk}

As matter is accreted from the torus onto the BH, magnetic flux is accumulated on the BH. This   triggers the BZ mechanism. It is expected that the value of the \Bf\ that can be accumulated on the BH, and thus the jet power, depend on the accretion rate. Let us next  estimate
the maximal luminosity of a magnetically driven jet.

The accretion rate from the disk is $\dot{M}_{d}=1.6M_{\rm d}/t_{\rm vis}$ \citep{ss73,mpq08}.
Wind from the disk surface can significantly reduce accretion rate at the inner parts of the disk.
Using the model of \cite{bb99} and following  \cite{mpq08}, the accretion rate onto the BH can be calculated as
\be
\frac{dM_{\rm BH}}{dt} =  \frac{1.6M_{\rm d}}{t_{\rm vis}}\left(\frac{r_{\rm ISCO}}{r_{\rm d}}\right)^{p} \left(\frac{(2p+1)t_{\rm vis}}{(2p+1)t_{\rm vis}+4.8t}\right)^{\frac{4(p+1)}{3}},
\label{eq:dmBHdt}
\ee
here $0<p<1$ is a non-dimensional parameter that describes wind intensity,
\be
r_{\rm d} = \left(\alpha_{\rm ss}^{2}h_{\rm d}^{4} GM_{\rm BH}t^{2}\right)^{1/3},
\label{eq:rdt}
\ee
here $h_{\rm d} = H/R$ is the relative thickness of the disk and $\alpha_{\rm ss}$ is the non-dimensional viscous parameter.
The radius of the innermost stable circular orbit (ISCO) is
\be
r_{\rm ISCO} = \frac{G M_{\rm BH} f(a_{\rm BH})}{c^2},
\label{eq:rISCO}
\ee
here $1<f(a_{\rm BH})<6$ is the non-dimensional radius of the last marginally stable orbit  \citep{bpt72}, which is a function of the BH spin parameter ``$a_{\rm BH}$''.

The accretion rate onto BH is
\be
\frac{dM_{\rm BH}}{dt} = \frac{0.16 (2p+1)^{\frac{4(p+1)}{3}} f(a_{\rm BH})^p M_{\rm d,-2} M_{\rm BH,0.5}^{\frac{2p}{3}} t_{\rm vis,-1}^{\frac{(2p+1)}{3}}}{60^p h_{\rm d,-0.5}^{\frac{4p}{3}}
\alpha_{\rm ss,-1}^{\frac{2p}{3}} ( (2p+1)t_{\rm vis,-1}+48t  )^{\frac{4(p+1)}{3}}} \quad M_{\odot}\mbox{ s}^{-1}.
\label{eq:dmBHdt_p}
\ee
For the typical parameters of the NS+NS mergers (see Section~\ref{sc:model}), if we assume a conservative value of $p=1/2$, the Equation~(\ref{eq:dmBHdt_p}) takes the form of
\be
\frac{dM_{\rm BH}}{dt} = 0.02\frac{ f(a_{\rm BH})^{1/2} M_{\rm d,-2} M_{\rm BH,0.5}^{1/3} t_{\rm vis,-1}^{2/3}}{  h_{\rm d,-0.5}^{2/3}
\alpha_{\rm ss,-1}^{1/3} ( t_{\rm vis,-1}+24t  )^{2}} \quad M_{\odot}\mbox{ s}^{-1}.
\label{eq:dmBHdt_p2}
\ee

The maximal jet power can be $L_{\rm BZ} = C(a_{\rm BH}) \dot{M}_{\rm BH} c^2$, here $C(a_{\rm BH})$ is the efficiency of the accretion coefficient, $0<C(a_{\rm BH})<$~few \citep[see, e.g.,][]{kb10,mtb12},
which can be combined with factor $f(a_{\rm BH})$ and, for a wide range of BH spin parameter ($a_{\rm BH}>0.3$), we obtain a simple approximation formula $C(a_{\rm BH}) f(a_{\rm BH})^{1/2} \approx a_{\rm BH}^{2.4}$.
If we adopt $a_{\rm BH}\approx 0.7$, we can estimate jet power as
\be
L_{\rm BZ} \sim 10^{49} M_{\rm d,-2} M_{\rm BH,0.5}^{1/3} t_{\rm vis,-1}^{2/3} t_{0.3}^{-2} \quad \mbox{erg s}^{-1}.
\label{eq:LBZ}
\ee
This equation is valid in the case $t \gg 0.4 t_{\rm vis}$.
So this electromagnetic luminosity corresponds to the strength of the magnetic field near the horizon of order \citep{bz77,bk08b}
\be
B_{\rm BH} \approx \left(\frac{24 L_{\rm BZ}}{r_{\rm g}^2 c}\right)^{1/2} \sim 2\times10^{14} \quad \mbox{G}.
\label{eq:Bbh}
\ee
This value, inferred from modeling the accretion torus, compares well with the one inferred from the soft $X$-ray component (Equation (\ref{BBH})).

\subsection{Optically Thick Pair-loaded Wind from the Disk}

The wind originates in a very optically thick and hot disk with $T\sim 1$ MeV. The optical thickness of the disk itself near the BH can be estimated from Equation~(\ref{eq:dmBHdt_p2}) $\tau_{\rm d}\sim3\times10^{11}t^{-2}$.
The wind is very optically thick,  even without pair creation the expected optical depth to Thomson scattering is $\tau_T \gg 1$,
\be
\tau_T \approx \frac{\sigma_T {M}_{\rm d}}{4 m_p  v_{\rm w}^2 t^2} \sim 5\times10^{10} {M}_{\rm d,-2} v_{\rm ex,-1}^{-2} t_{0.3}^{-2},
\ee
here we assume uniform density of the disk wind $\rho_{\rm w} = M_{\rm d}/4 r_{\rm w}^3$. This result depends on the wind parameter $p$, which is neglected here.
A large number of electron-positron pairs will be created near the disk. However, the $e^\pm$ pairs do not affect wind dynamics since baryon density exceeds the pair density by many orders of magnitude,
\ba &&
\frac{\rho_{\rm w}} { m_e n_\pm} = \frac{\sqrt{\pi/2}}{4}\frac{e^{r/(r_0 \theta_{T,0})} }{m_e v_{\rm w} r^{1/2} r_0^{3/2}\theta_{T,0}^{3/2}} \gg 1
\nn &&
 n_\pm \lambda_C^3 =  \frac{\sqrt{2}}{\pi ^{3/2}}e^{-\frac{1}{\theta _T}} \theta _T^{3/2},
\ea
where  $\lambda_C=\hbar/(m_e c) $ is the electron Compton length and
 we assumed the wind temperature is $T =T_0 (r_0/r)$, { normalized $\theta_{T} = T/ m_e c^2 $ } (and similar for $T_0$) and used the equilibrium  thermal pair density $ n_\pm$
  \citep{1982ApJ...253..842L,1982ApJ...258..335S}.

The wind will not produce any appreciable EM signatures due to steep adiabatic cooling.
It is expected that the decay of radioactive elements contributes to wind heating, which together with far-flung tidal tails produce  the kilonova emission
\citep{1998ApJ...507L..59L, 2010MNRAS.406.2650M}.

\subsection{Jet Propagation within the Wind}
\label{propagate}

Next, we consider the propagation of the BZ jet through the disk-generated wind. Since it takes time to accumulate the magnetic flux on the BH and to trigger the BZ mechanism, the jet will propagate through the pre-existing wind. Let the wind have constant properties (we neglect the fact that at early times the wind has higher $\dot{M}$; we might expect that this will be partially compensated by the initially higher $v_{\rm w}$).
Consider a jet of power $L_j$ that is confined within a solid angle $\Delta \Omega\approx \pi\theta^2$. The head of the jet located at $r_h$ propagates with speed $v_{\rm jh}$ and  according to
\citep[in the Kompaneets approximation,][]{Komp}
\be
P_{\rm jh}=\frac{L_j}{\pi \theta^2 r_h^2 c} = \rho_{\rm w} (v_{\rm jh} -  v_{\rm w})^2.
\label{eq:Pjh}
\ee
where $P_{\rm jh}$ is the pressure created by the jet.
Let us assume that the disk provides an outflow
\be
\dot{M}_{\rm w,disk}= 4 \pi v_{\rm w} \rho_{\rm w} r^2.
\label{rhow}
\ee
Using wind density from (\ref{rhow}),  the jet head propagates according to \cite[see ][]{bnp11}
\be
v_{\rm jh} = v_{\rm w} + \sqrt{\frac{4 L_j v_{\rm w}}{\theta^2\dot{M}_{\rm w,disk} c}}
\label{eq:vj}
\ee
or
\be
v_{\rm jh} =v_{\rm w} + 5 \times 10^9 L_{j,49}^{1/2}v_{\rm w,-1}^{1/2}\dot{M}_{\rm w,disk,-2}^{-1/2}  {\rm cm}\;{\rm s}^{-1}.
\ee

Thus, if the jet propagates with nearly constant velocity, then
the jet propagation time is
\be
t_{\rm jb} = \frac{v_{\rm w}t}{v_{\rm jh}}\sim \frac{t_{\rm delay}}{3},
\label{eq:jpt}
\ee
at the second half of this equation, we assume  $v_{\rm w}=0.1c=3\times10^9$~cm s$^{-1}$ and $v_{\rm jh}=5\times 10^9$~cm s$^{-1} \approx 2 v_{\rm w}$.
Moreover, if we again assume $v_{\rm w}=0.1c$, this delay time is in good agreement with results of statistical analysis of short GRBs \citep{mp17}.
Importantly, the head propagates  within the wind with  non-relativistic velocity.

 If the speed of jet's head shock in the wind rest frame $v_{\rm s} = v_{\rm jh} - v_{\rm w}\ge v_{\rm w}$  is less than the wind speed, then the jet cocoon would have an approximately spherical shape in the expanding disk wind. And, in this case, after the shock breakout the head shock is not intense and could not form relativistic outflow and form a sufficiently large impact region at the boundary of the wind with the radius of $r_{\rm s}\sim r_{\rm w}$.
So, comparing $v_{\rm s}$ and $v_{\rm w}$ from Equations~(\ref{eq:vj}) and (\ref{eq:thetaj}), we can derive a restriction on the jet power as
\be
L_j \ge \frac{S\dot{M}_{\rm w,disk} c}{4\pi v_{\rm w} t_{\rm delay}^2}\sim 3\times 10^{48} \dot{M}_{\rm w,disk,-2} v^{-1}_{w,-1} \mbox{ erg s}^{-1}.
\label{eq:jcm}
\ee

\begin{figure*}
\includegraphics[width=88mm,angle=-0]{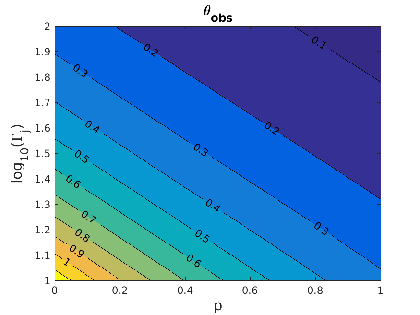}
\includegraphics[width=88mm,angle=-0]{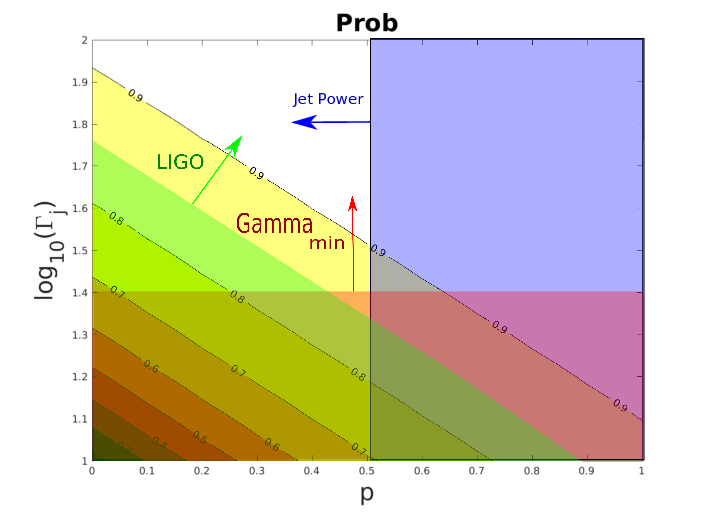}
\caption{ Observational angle (left) and probability to observe the transient (right) depends on wind parameter "$p$" and logarithm of the jet Lorentz factor $\Gamma_{\rm jet}$.
In the right plot, we put limitation on the wind "$p$" parameter from Equations~(\ref{eq:jcm}), (\ref{Lj}),
and (\ref{eq:dmBHdt_p}) as purple area. The green area is following from LIGO observations.}
\label{fig:theta}
\end{figure*}

\subsection{Jet Lorentz Factor and Its Dynamics}
\label{sc:ja}

Let us estimate the Lorentz factor of the magnetically driven jet before the breakout from the wind.
Magnetically driven jets can be accelerated up to very high Lorentz factors in the linear regime \citep{bn06,bk08,kvkb09}.
In the case of the parabolic shape of the jet, we can estimate the jet Lorentz factor on the border expanding envelop as $\Gamma_{\rm j} = (r_{\rm S}/r_{\rm lc})^{1/2}$,
here the cylindrical radius of the jet $r_{\rm S}=6\times 10^8$~cm and light cylinder radius $r_{\rm lc}$. In a case of fast spinning BH, $r_{\rm lc} = 4r_{\rm g}/a_{\rm BH} \approx 2\times10^6$~cm,
for model parameters adopted in Section \ref{sc:model} this corresponds to $\Gamma_{\rm j}\approx 17$.
We can treat this value as the minimal Lorentz factor of the jet. After the breakout, the jet can have additional boosting due to the formation of a strong rarefaction wave on its outer border
\citep[see ][]{kvk10,tnm10,lyut11,kbg17}.  The additional opening angle, which is formed by sideways jet expansion, can be estimated as $\Delta\theta\approx1/\Gamma_{\rm j}\approx 0.06$ \citep{2003ApJ...597..998L}.
So, the total opening angle of the jet can be estimated as $\theta_{\rm tot}= \theta + \Delta\theta \approx 0.15$.

\subsection{The Second Thermal Peak}

We can estimate the properties of the shock-heated wind flow at the point of breakout. Before the breakout, the
 BZ jet  drives a shock in the wind. This shock is strongly radiation-dominated, {as we demonstrate next \citep[see][for a discussion of radiation-dominated shocks]{2010ApJ...716..781K,ilsn17}.
 Equating post-shock plasma energy density to that of radiation, a shock becomes  radiation-dominated for
 the post-shock (ion) temperatures $T_r$ that satisfies
 \be
 T_r > \left(\frac{15 n}{\pi^2 }\right)^{1/3} \frac{c \hbar}{k},
 \ee
 where $n$ is the pre-shock number density, and $k$ is the Boltzmann constant. From the condition $kT=m_p v_r^2$, the required shock velocity is then
 \be
 v _r\geq  \sqrt{ \frac{\hbar c}{m_p }} \left(\frac{15 n }{\pi^2}\right)^{1/6}.
 \ee
 Comparing this with the relative velocity of the jet head with respect to the wind, (\ref{eq:vj}),
 \be
 \frac{v_{\rm jh} - v_{\rm w}}{v_r} =\frac{2^{4/3}\pi^{1/2}}{15^{1/6}} \frac{m_{\rm p}^{2/3}}{c \hbar^{1/2}} \frac{ L_{j}^{1/2} v_{\rm w} t^{1/3}}{  \theta \dot{M}_{d,wind}^{2/3} }
 \label{eq:vrel}
 \ee
 or
 \be
 \frac{v_{\rm jh} - v_{\rm w}}{v_r} = 70 \frac{ L_{j,49}^{1/2} v_{\rm w,-1} t_{0.3}^{1/3}}{  \theta \dot{M}_{d,wind,-2}^{2/3} }.
 \label{eq:vreln}
 \ee
The speed of the shock wave significantly exceeds the critical one, so  we conclude that the jet-driven shock is radiatively dominated close to the breakout point.}

In  radiation-dominated shocks the post-shock temperature $T_{\rm s}$ is determined by the condition
$$
\frac{4 \sigma_{SB}}{c} T_{\rm s}^4 \sim \rho_{\rm w} (v_{\rm jh} - v_{\rm w})^2 = \frac{L_{\rm j}}{\pi (v_{\rm w} t \theta)^2 c}
$$
\be
T_{\rm s} =  \frac{L_j^{1/4}}{ (4\pi\sigma_{\rm SB})^{1/4} \sqrt{ v_{\rm w} t \theta}  }.
\label{eq:Ts}
\ee
(Note that this estimate has a weak dependence on the properties of the jet (power and opening angle), is  independent of the assumed mass loss rate of the disk, and only mildly dependent on the wind velocity.)

Substituting jet power from Equation~(\ref{eq:LBZ}), we get
\be
T_{\rm s} = 40 \frac{M_{\rm d,-2}^{1/4}}{\theta_{-1}^{1/2}v_{\rm w,-1}^{1/2} t_{0.3}} \mbox{ keV},
\ee
where time $t$ is time of shock breakout. Estimate $T_{\rm s}$ is the upper limit since part of the energy incoming into the shock will be converted  into bulk motion and into the production of pairs  --
these effects will reduce somewhat  the post-shock temperature. At $t\sim $ a few seconds, and given our order-of-magnitude approach,  this estimate is very close to the observed temperature of the second prompt peak.

\subsection{Estimate of the Viewing Angle}
\label{ssc:va}

In the present model, after the breakout, the collimated jet accelerates to highly relativistic velocities, and thus becomes highly debeamed and not observable.
Let us estimate the constraints  on the  viewing angle.

The expected observed jet power of a  relativistically boosted jet can be estimated  as \citep[see more details in ][]{smmp97}
\be
L_{\rm obs} = \frac{\delta^3}{\Gamma_{\rm jet}}L_{\rm jet},
\label{eq:Lboost}
\ee
here $\Gamma_{\rm jet}$ is the Lorentz factor of the jet,  $\delta = 1/[\Gamma_{\rm jet}(1-v\cos\theta/c)]$ is the Doppler factor of the jet, $v$ is the velocity of the jet, $\theta$
is the angle between observer and jet axis, and $L_{\rm jet}=\chi L_{\rm BZ}$ is the electromagnetic jet luminosity with efficiency $\chi$. In the case of small angles
($|1- \cos \theta|\ll 1$, but $\theta$ can be $\gg 1/\Gamma_{\rm jet}$), we can write $\Gamma_{\rm jet}/\delta = (\alpha^2+1)/2$ \citep{gub10,abk17},
here $\alpha = \theta \Gamma_{\rm jet}$ and Equation~(\ref{eq:Lboost}) take the form
\be
\alpha = \left[\left(\frac{8\Gamma_{\rm jet}^2\chi L_{\rm BZ} }{L_{\rm obs}}\right)^{1/3}-1\right]^{1/2}.
\label{eq:alpha}
\ee
In the case of large viewing angles ($\alpha\gg 1 $), the Equation~(\ref{eq:alpha}) can be writen as
\be
\theta_{\rm side}=\left(\frac{8\chi L_{\rm BZ} }{\Gamma_{\rm jet}^4 L_{\rm obs}}\right)^{1/6}.
\label{eq:thmin}
\ee
So, taking the jet opening angle from Section~\ref{sc:ja}, the minimal viewing angle can be estimated as
\be
\theta_{\rm obs} > \theta_{\rm tot} + \left(\frac{8\chi L_{\rm BZ} }{\Gamma_{\rm jet}^4 L_{\rm obs}}\right)^{1/6}.
\label{eq:theta}
\ee

Substituting parameters of the  GRB~170817A to Equation~(\ref{eq:theta}), we can estimate the minimal viewing angle as
\be
\theta_{\rm obs} > 0.15 + 0.3 \left(\frac{\chi M_{\rm d0,-2} M_{\rm BH,0.5}^{1/3} t_{\rm vis,-1}^{2/3} }{\Gamma_{\rm jet,15}^4 L_{\rm obs,47.2} \alpha_{\rm ss,-1} h_{\rm d,-0.5} t_{0.3}^{2}}\right)^{1/6}.
\label{eq:thetar}
\ee
The probability to not observe such a burst is $Prob \approx 1 -  \theta_{\rm obs}^2/2 \approx 1 - 0.1 \approx 0.9$.
We should note that our limitation of observational angle is close to the value $\theta_{\rm GW}=0.54$ obtained  from BNS coalescence \citep{MMA2017}.

Combining Equations (\ref{eq:dmBHdt_p}) and  (\ref{eq:alpha}), we derive a minimal observational angle and probability to observe the transient and present them in Figure~\ref{fig:theta}.

\section{Conclusions}
\label{sc:dis}

In this paper, we provide the results of the coordinated optical and radio search and optical observations of the LIGO/\textit{Fermi} GBM event GW170817/GRB~170817A and propose a new theoretical model to explain the prompt gamma emission.

An optical transient of $\sim 19^m$ was detected by the Chilescope observatory. The properties of the optical transient match the kilonova activity \citep{2017Natur.551...75S}. We also provide upper limits on the possible short radio transient at 110
MHz associated with the GW170817/GRB~170817A at the trigger time.

We discuss the prompt gamma-ray emission, consisting of a hard gamma-ray pulse followed by a soft tail, each delayed by $\sim 2$ s with respect to the LIGO trigger.
The appearance of the thermal component at the end of the burst is unusual for short GRBs.   Both the hard pulse and the whole burst do not fit the Amati relation for short GRBs. This is especially true for the first hard pulse.

We then developed a  theoretical model, which must explain  (i) the delay between the LIGO and \textit{Fermi} triggers; (ii) the two-component nature of the prompt emission;  (iii)  the absence of an afterglow.

{\it Delay between LIGO and Fermi signals.} The delay between the LIGO  and the EM signals depends both on the delayed switching-on of the BZ-powered jet and the jet
propagation through the dense expanding envelope \citep[see Equation~(\ref{eq:jpt}) and ][]{gnp17}. The neutrino mechanism of jet launching is very sensitive to
the disk temperature and should be more effective right after the disk formation; at this time, the accretion rate is also maximal. On the other hand,
the  BZ mechanism is less sensitive to accretion rate and needs sufficient magnetic flux accumulated on the BH, \cite[see, e.g., ][]{KB09}. Also,
an operation of the BZ mechanism requires magnetically dominated plasma - the cleaning of polar regions of the BH can
take about  1~s \citep{bk10}. We suggest the observed delay results from both the delay of activation of the BZ jet and the jet propagation through expanding envelope.

{\it Two components of prompt emission.}
We suggest in our model that the two components come from, first, highly relativistic, nearly isotropic breakout of a magnetized jet from the confining wind,
and, second, from the wind material heated by the breaking-out jet. After activation of the BZ process,  the jet head propagates through wind matter
with subrelativistic velocity.
The initially strongly magnetically dominated jet partially transfers its magnetic energy into internal energy, as discussed in  Section \ref{spike}.
During the jet breakout from the confining wind, the jet expands quasi-isotropically (see Figure~\ref{fig:SK}).
The expanding matter  accelerates up to the high Lorentz factor (see Equation~(\ref{eq:gjmin})) at the radius of $r_{\rm ha} \approx
\Gamma_{\rm j, min} r_{\rm S} \approx 10^{10}$~cm,  where the cloud becomes transparent for radiation.
The minimal variability time $\sim r_{\rm s}/c \sim 0.02$~s does not contradict the duration of the first hard pulse.
The second soft gamma-ray component is formed by thermal emission of shocked stellar wind matter. This interpretation allows us to put self-consistent restrictions on
the cylindrical radius of the jet $r_{\rm S} = 6\times10^8$~cm, power of the jet $L_{\rm jet} \sim 10^{49}\mbox{ erg s}^{-1}$   and jet penetration time through the wind envelope.
Also, we estimate jet power which can be supplied by the magnetically driven jet supported by the accretion disk with the wind (see Equations~(\ref{eq:jcm}), (\ref{Lj}), and (\ref{eq:LBZ})).

{\it Absence of early afterglows.}
We suggest that the absence of the afterglow can be explained by the observer off-axis to the jet. After the jet breaks out, a standard jet acceleration will be recovered  - the BZ flow accelerates to $\Gamma \geq 17$
at the edge of the wind and after that jet boosts its acceleration. Later, the jet becomes  causally disconnected and propagates ballistically. This boost in acceleration will suppress the visibility of the BZ jet, and the corresponding forward shock. Importantly, the luminosity of an afterglow component of GRB~170817A is less by factors at least $30$ in $r$-filter and $130$ in $J$-filter than the afterglow of GRB~130603B at $\sim$11 hr after burst onset (see Figure~\ref{fig:kilonova+AG}).

\section*{Acknowledgments}

A.S.P., P.Yu.M., A.A.V., and E.D.M. are grateful to RFBR grants 17-02-01388 and 17-52-80139 for partial support.
This work was supported by NSF grant AST-1306672, DoE grant DE-SC0016369, and NASA grant 80NSSC17K0757.
We acknowledge the excellent help in obtaining Chilescope data provided by Sergei Pogrebisskiy and Ivan Rubtsov. We are grateful to the anonymous referee for the thorough reading of the article and for useful comments that contributed to the improvement of the article.



\begin{thebibliography}{130}
\expandafter\ifx\csname natexlab\endcsname\relax\def\natexlab#1{#1}\fi

\bibitem[{{Aharonian} {et~al.}(2017){Aharonian}, {Barkov}, \&
  {Khangulyan}}]{abk17}
{Aharonian}, F.~A., {Barkov}, M.~V., \& {Khangulyan}, D. 2017, \apj, 841, 61

\bibitem[{{Allam} {et~al.}(2017){Allam}, {Annis}, {Berger}, \&
  et~al.}]{decam-gcn}
{Allam}, S., {Annis}, J., {Berger}, E., \& et~al. 2017, GCN, 21530 LVC

\bibitem[{{Amati}(2010)}]{amati_10}
{Amati}, L. 2010, ArXiv:1002.2232

\bibitem[{{Arcavi} {et~al.}(2017){Arcavi}, {Howell}, {McCully}, \&
  et~al.}]{lco-gcn}
{Arcavi}, I., {Howell}, D.~A., {McCully}, C., \& et~al. 2017, GCN, 21538 LVC

\bibitem[{{Baiotti} \& {Rezzolla}(2017)}]{bairez17}
{Baiotti}, L., \& {Rezzolla}, L. 2017, Reports on Progress in Physics, 80,
  096901

\bibitem[{{Bardeen} {et~al.}(1972){Bardeen}, {Press}, \& {Teukolsky}}]{bpt72}
{Bardeen}, J.~M., {Press}, W.~H., \& {Teukolsky}, S.~A. 1972, \apj, 178, 347

\bibitem[{{Barkov}(2008)}]{b08}
{Barkov}, M.~V. 2008, in American Institute of Physics Conference Series, Vol.
  1054, American Institute of Physics Conference Series, ed. M.~{Axelsson},
  79--85

\bibitem[{{Barkov} \& {Baushev}(2011)}]{bb11}
{Barkov}, M.~V., \& {Baushev}, A.~N. 2011, \na, 16, 46

\bibitem[{{Barkov} \& {Komissarov}(2008{\natexlab{a}})}]{bk08}
{Barkov}, M.~V., \& {Komissarov}, S.~S. 2008{\natexlab{a}}, International
  Journal of Modern Physics D, 17, 1669

\bibitem[{{Barkov} \& {Komissarov}(2008{\natexlab{b}})}]{bk08b}
---. 2008{\natexlab{b}}, \mnras, 385, L28

\bibitem[{{Barkov} \& {Komissarov}(2010)}]{bk10}
---. 2010, \mnras, 401, 1644

\bibitem[{{Barkov} \& {Pozanenko}(2011)}]{bp11}
{Barkov}, M.~V., \& {Pozanenko}, A.~S. 2011, \mnras, 417, 2161

\bibitem[{{Barnes} \& {Kasen}(2013)}]{bk13}
{Barnes}, J., \& {Kasen}, D. 2013, \apj, 775, 18

\bibitem[{{Barnes} {et~al.}(2016){Barnes}, {Kasen}, {Wu}, \&
  {Mart{\'{\i}}nez-Pinedo}}]{bkw16}
{Barnes}, J., {Kasen}, D., {Wu}, M.-R., \& {Mart{\'{\i}}nez-Pinedo}, G. 2016,
  \apj, 829, 110

\bibitem[{{Bauswein} {et~al.}(2013){Bauswein}, {Goriely}, \& {Janka}}]{bgj13}
{Bauswein}, A., {Goriely}, S., \& {Janka}, H.-T. 2013, \apj, 773, 78

\bibitem[{{Beloborodov}(2003)}]{2003ApJ...588..931B}
{Beloborodov}, A.~M. 2003, \apj, 588, 931

\bibitem[{{Beloborodov}(2010)}]{2010MNRAS.407.1033B}
---. 2010, \mnras, 407, 1033

\bibitem[{{Beloborodov}(2017)}]{2017ApJ...838..125B}
---. 2017, \apj, 838, 125

\bibitem[{{Beloborodov} \& {M{\'e}sz{\'a}ros}(2017)}]{2017SSRv..207...87B}
{Beloborodov}, A.~M., \& {M{\'e}sz{\'a}ros}, P. 2017, \ssr, 207, 87

\bibitem[{{Beskin} \& {Nokhrina}(2006)}]{bn06}
{Beskin}, V.~S., \& {Nokhrina}, E.~E. 2006, \mnras, 367, 375

\bibitem[{{Birkl} {et~al.}(2007){Birkl}, {Aloy}, {Janka}, \&
  {M{\"u}ller}}]{bir07}
{Birkl}, R., {Aloy}, M.~A., {Janka}, H., \& {M{\"u}ller}, E. 2007, {\aa}p, 463,
  51

\bibitem[{{Bisnovatyi-Kogan}(1970)}]{bk70}
{Bisnovatyi-Kogan}, G.~S. 1970, \azh, 47, 813

\bibitem[{{Blandford} \& {Begelman}(1999)}]{bb99}
{Blandford}, R.~D., \& {Begelman}, M.~C. 1999, \mnras, 303, L1

\bibitem[{{Blandford} \& {Znajek}(1977)}]{bz77}
{Blandford}, R.~D., \& {Znajek}, R.~L. 1977, \mnras, 179, 433

\bibitem[{{Blinnikov} {et~al.}(1984){Blinnikov}, {Novikov}, {Perevodchikova},
  \& {Polnarev}}]{bnpp84}
{Blinnikov}, S.~I., {Novikov}, I.~D., {Perevodchikova}, T.~V., \& {Polnarev},
  A.~G. 1984, Pis ma Astronomicheskii Zhurnal, 10, 422

\bibitem[{{Bromberg} {et~al.}(2011){Bromberg}, {Nakar}, {Piran}, \&
  {Sari}}]{bnp11}
{Bromberg}, O., {Nakar}, E., {Piran}, T., \& {Sari}, R. 2011, \apj, 740, 100

\bibitem[{{Brown} {et~al.}(2013){Brown}, {Baliber}, {Bianco}, \&
  et~al.}]{lco-about}
{Brown}, T.~M., {Baliber}, N., {Bianco}, F.~B., \& et~al. 2013, \pasp, 125,
  1031

\bibitem[{{Cash}(1979)}]{cash79}
{Cash}, W. 1979, \apj, 228, 939

\bibitem[{{Connaughton} {et~al.}(2017){Connaughton}, {Blackburn}, {Briggs},
  {Broida}, \& {Burns}}]{gcn21506}
{Connaughton}, V., {Blackburn}, L., {Briggs}, M.~S., {Broida}, J., \& {Burns},
  E. 2017, GCN, 21506 LVC

\bibitem[{{Coulter} {et~al.}(2017){Coulter}, {Kilpatrick}, {Siebert}, {Foley},
  {Shappee}, {Drout}, {Simon}, \& {Piro}}]{Swope-gcn}
{Coulter}, D.~A., {Kilpatrick}, C.~D., {Siebert}, M.~R., {Foley}, R.~J.,
  {Shappee}, B.~J., {Drout}, M.~R., {Simon}, J.~S., \& {Piro}, A.~L. 2017, GCN, 21529 LVC

\bibitem[{{Dalya} {et~al.}(2016){Dalya}, {Frei}, {Galgoczi}, {Raffai}, \& {de
  Souza}}]{galaxies-list}
{Dalya}, G., {Frei}, Z., {Galgoczi}, G., {Raffai}, P., \& {de Souza}, R.~S.
  2016, VizieR Online Data Catalog, 7275

\bibitem[{{Derishev} {et~al.}(1999){Derishev}, {Kocharovsky}, \&
  {Kocharovsky}}]{1999ApJ...521..640D}
{Derishev}, E.~V., {Kocharovsky}, V.~V., \& {Kocharovsky}, V.~V. 1999, \apj,
  521, 640

\bibitem[{{Eichler} \& {Levinson}(2004)}]{eic04}
{Eichler}, D., \& {Levinson}, A. 2004, \apjl, 614, L13

\bibitem[{{Eichler} {et~al.}(1989{\natexlab{a}}){Eichler}, {Livio}, {Piran}, \&
  {Schramm}}]{1989Natur.340..126E}
{Eichler}, D., {Livio}, M., {Piran}, T., \& {Schramm}, D.~N.
  1989{\natexlab{a}}, \nat, 340, 126

\bibitem[{{Eichler} {et~al.}(1989{\natexlab{b}}){Eichler}, {Livio}, {Piran}, \&
  {Schramm}}]{elps89}
---. 1989{\natexlab{b}}, \nat, 340, 126

\bibitem[{{Ensman} \& {Burrows}(1992)}]{1992ApJ...393..742E}
{Ensman}, L., \& {Burrows}, A. 1992, \apj, 393, 742

\bibitem[{{Fenimore} {et~al.}(1995){Fenimore}, {in 't Zand}, {Norris},
  {Bonnell}, \& {Nemiroff}}]{fen95}
{Fenimore}, E.~E., {in 't Zand}, J.~J.~M., {Norris}, J.~P., {Bonnell}, J.~T.,
  \& {Nemiroff}, R.~J. 1995, \apjl, 448, L101

\bibitem[{{Giannios} {et~al.}(2010){Giannios}, {Uzdensky}, \&
  {Begelman}}]{gub10}
{Giannios}, D., {Uzdensky}, D.~A., \& {Begelman}, M.~C. 2010, \mnras, 402, 1649

\bibitem[{{Goldstein} {et~al.}(2017{\natexlab{a}}){Goldstein}, {Veres},
  {Burns}, \& et~al.}]{gol17}
{Goldstein}, A., {Veres}, P., {Burns}, E., \& et~al. 2017{\natexlab{a}}, \apjl,
  848, L14

\bibitem[{{Goldstein} {et~al.}(2017{\natexlab{b}}){Goldstein}, {Veres}, {von
  Kienlin}, {Blackburn}, \& {Briggs}}]{gcn21528}
{Goldstein}, A., {Veres}, P., {von Kienlin}, A., {Blackburn}, L., \& {Briggs},
  M.~S. 2017{\natexlab{b}}, GCN, 21528 LVC

\bibitem[{{Goodman}(1986)}]{goodman86}
{Goodman}, J. 1986, \apjl, 308, L47

\bibitem[{{Gottlieb} {et~al.}(2018){Gottlieb}, {Nakar}, \& {Piran}}]{gnp17}
{Gottlieb}, O., {Nakar}, E., \& {Piran}, T. 2018, \mnras, 473, 576

\bibitem[{{Gottlieb} {et~al.}(2017){Gottlieb}, {Nakar}, {Piran}, \&
  {Hotokezaka}}]{gnph17}
{Gottlieb}, O., {Nakar}, E., {Piran}, T., \& {Hotokezaka}, K. 2017, ArXiv
 :astro-ph/1710.05896

\bibitem[{{Gralla} \& {Jacobson}(2014)}]{2014MNRAS.445.2500G}
{Gralla}, S.~E., \& {Jacobson}, T. 2014, \mnras, 445, 2500

\bibitem[{{Grossman} {et~al.}(2014){Grossman}, {Korobkin}, {Rosswog}, \&
  {Piran}}]{gkr14}
{Grossman}, D., {Korobkin}, O., {Rosswog}, S., \& {Piran}, T. 2014, \mnras,
  439, 757

\bibitem[{{Gruber} {et~al.}(2014){Gruber}, {Goldstein}, {Weller von Ahlefeld},
  \& et~al.}]{gruber14}
{Gruber}, D., {Goldstein}, A., {Weller von Ahlefeld}, V., \& et~al. 2014,
  \apjs, 211, 12

\bibitem[{{Gruzinov}(1999)}]{Gruzinov99}
{Gruzinov}, A. 1999, ArXiv:9902288

\bibitem[{{Ito} {et~al.}(2017){Ito}, {Levinson}, {Stern}, \&
  {Nagataki}}]{ilsn17}
{Ito}, H., {Levinson}, A., {Stern}, B.~E., \& {Nagataki}, S. 2017, ArXiv
  e-prints

\bibitem[{{Johnson} \& {McKee}(1971)}]{JohnsonMcKee}
{Johnson}, M.~H., \& {McKee}, C.~F. 1971, \prd, 3, 858

\bibitem[{{Jones} {et~al.}(2009){Jones}, {Read}, {Saunders}, \& {et
  al.}}]{2009MNRAS.399..683J}
{Jones}, D.~H., {Read}, M.~A., {Saunders}, W., \& {et al.} 2009, \mnras, 399,
  683

\bibitem[{{Kasen} {et~al.}(2013){Kasen}, {Badnell}, \& {Barnes}}]{kbb13}
{Kasen}, D., {Badnell}, N.~R., \& {Barnes}, J. 2013, \apj, 774, 25

\bibitem[{{Kathirgamaraju} {et~al.}(2017){Kathirgamaraju}, {Barniol Duran}, \&
  {Giannios}}]{kbg17}
{Kathirgamaraju}, A., {Barniol Duran}, R., \& {Giannios}, D. 2017,
  ArXiv:1708.07488

\bibitem[{{Katz} {et~al.}(2010){Katz}, {Budnik}, \&
  {Waxman}}]{2010ApJ...716..781K}
{Katz}, B., {Budnik}, R., \& {Waxman}, E. 2010, \apj, 716, 781

\bibitem[{{Kennel} \& {Coroniti}(1984)}]{kc84}
{Kennel}, C.~F., \& {Coroniti}, F.~V. 1984, \apj, 283, 694

\bibitem[{{Komissarov}(2011)}]{2011MNRAS.418L..94K}
{Komissarov}, S.~S. 2011, \mnras, 418, L94

\bibitem[{{Komissarov} \& {Barkov}(2009)}]{KB09}
{Komissarov}, S.~S., \& {Barkov}, M.~V. 2009, \mnras, 397, 1153

\bibitem[{{Komissarov} \& {Barkov}(2010)}]{kb10}
---. 2010, \mnras, 402, L25

\bibitem[{{Komissarov} {et~al.}(2010){Komissarov}, {Vlahakis}, \&
  {K{\"o}nigl}}]{kvk10}
{Komissarov}, S.~S., {Vlahakis}, N., \& {K{\"o}nigl}, A. 2010, \mnras, 407, 17

\bibitem[{{Komissarov} {et~al.}(2009){Komissarov}, {Vlahakis}, {K{\"o}nigl}, \&
  {Barkov}}]{kvkb09}
{Komissarov}, S.~S., {Vlahakis}, N., {K{\"o}nigl}, A., \& {Barkov}, M.~V. 2009,
  \mnras, 394, 1182

\bibitem[{{Kompaneets}(1957)}]{komp56}
{Kompaneets}, A.~S. 1957, Sov. Phys. JETP, 4, 730

\bibitem[{{Kompaneets}(1960)}]{Komp}
---. 1960, Soviet Physics Doklady, 5, 46

\bibitem[{{Lazzati} {et~al.}(2017){Lazzati}, {L{\'o}pez-C{\'a}mara},
  {Cantiello}, {Morsony}, {Perna}, \& {Workman}}]{llc17}
{Lazzati}, D., {L{\'o}pez-C{\'a}mara}, D., {Cantiello}, M., {Morsony}, B.~J.,
  {Perna}, R., \& {Workman}, J.~C. 2017, \apjl, 848, L6

\bibitem[{{LeBlanc} \& {Wilson}(1970)}]{LW70}
{LeBlanc}, J.~M., \& {Wilson}, J.~R. 1970, \apj, 161, 541

\bibitem[{{Levinson} \& {Eichler}(2005)}]{lev05}
{Levinson}, A., \& {Eichler}, D. 2005, \apjl, 629, L13

\bibitem[{{Li} \& {Paczy{\'n}ski}(1998)}]{1998ApJ...507L..59L}
{Li}, L.-X., \& {Paczy{\'n}ski}, B. 1998, \apjl, 507, L59

\bibitem[{{Lightman}(1982)}]{1982ApJ...253..842L}
{Lightman}, A.~P. 1982, \apj, 253, 842

\bibitem[{{LIGO Scientific Collaboration} \& {Virgo
  Collaboration}(2017)}]{GW-trigger}
{LIGO Scientific Collaboration}, \& {Virgo Collaboration}. 2017, GCN, 21505 LVC

\bibitem[{{LIGO Scientific Collaboration} {et~al.}(2017){LIGO Scientific
  Collaboration}, {VIRGO Collaboration}, {Partner Astronomy Groups}, {Abbott},
  \& et~al.}]{MMA2017}
{LIGO Scientific Collaboration}, {VIRGO Collaboration}, {Partner Astronomy
  Groups}, {Abbott}, \& et~al. 2017, \apjl, 848, L13

\bibitem[{{Lipunov} {et~al.}(2010){Lipunov}, {Kornilov}, {Gorbovskoy}, \&
  et~al.}]{master}
{Lipunov}, V., {Kornilov}, V., {Gorbovskoy}, E., \& et~al. 2010, Advances in
  Astronomy, 2010, 349171

\bibitem[{{Lipunov} {et~al.}(2017){Lipunov}, {Gorbovskoy}, {Kornilov}, \&
  et~al.}]{master-gcn}
{Lipunov}, V.~M., {Gorbovskoy}, E., {Kornilov}, V.~G., \& et~al. 2017, GCN, 21546 LVC

\bibitem[{{Lithwick} \& {Sari}(2001)}]{2001ApJ...555..540L}
{Lithwick}, Y., \& {Sari}, R. 2001, \apj, 555, 540

\bibitem[{{Lyutikov}(2006)}]{2006NJPh....8..119L}
{Lyutikov}, M. 2006, New Journal of Physics, 8, 119

\bibitem[{{Lyutikov}(2010)}]{2010PhRvE..82e6305L}
---. 2010, \pre, 82, 056305

\bibitem[{{Lyutikov}(2011{\natexlab{a}})}]{lyut11}
---. 2011{\natexlab{a}}, \mnras, 411, 422

\bibitem[{{Lyutikov}(2011{\natexlab{b}})}]{2011PhRvD..83l4035L}
---. 2011{\natexlab{b}}, \prd, 83, 124035

\bibitem[{{Lyutikov}(2011{\natexlab{c}})}]{2011PhRvD..83f4001L}
---. 2011{\natexlab{c}}, \prd, 83, 064001

\bibitem[{{Lyutikov} \& {Hadden}(2012)}]{2012PhRvE..85b6401L}
{Lyutikov}, M., \& {Hadden}, S. 2012, \pre, 85, 026401

\bibitem[Lyutikov et al.(2003)]{2003ApJ...597..998L} Lyutikov, M., Pariev, V.~I., \& Blandford, R.~D.\ 2003, \apj, 597, 998

\bibitem[{{Lyutikov} \& {Usov}(2000)}]{LyutikovUsov}
{Lyutikov}, M., \& {Usov}, V.~V. 2000, \apjl, 543, L129

\bibitem[{{McKinney} {et~al.}(2012){McKinney}, {Tchekhovskoy}, \&
  {Blandford}}]{mtb12}
{McKinney}, J.~C., {Tchekhovskoy}, A., \& {Blandford}, R.~D. 2012, \mnras, 423,
  3083

\bibitem[{{Metzger} {et~al.}(2010){Metzger}, {Mart{\'{\i}}nez-Pinedo},
  {Darbha}, \& et~al.}]{2010MNRAS.406.2650M}
{Metzger}, B.~D., {Mart{\'{\i}}nez-Pinedo}, G., {Darbha}, S., \& et~al. 2010,
  \mnras, 406, 2650

\bibitem[{{Metzger} {et~al.}(2008{\natexlab{a}}){Metzger}, {Piro}, \&
  {Quataert}}]{mpq08}
{Metzger}, B.~D., {Piro}, A.~L., \& {Quataert}, E. 2008{\natexlab{a}}, \mnras,
  390, 781

\bibitem[{{Metzger} {et~al.}(2008{\natexlab{b}}){Metzger}, {Quataert}, \&
  {Thompson}}]{2008MNRAS.385.1455M}
{Metzger}, B.~D., {Quataert}, E., \& {Thompson}, T.~A. 2008{\natexlab{b}},
  \mnras, 385, 1455

\bibitem[{{Minaev} \& {Pozanenko}(2017)}]{min17}
{Minaev}, P.~Yu., \& {Pozanenko}, A.~S. 2017, Astronomy Letters, 43, 1

\bibitem[{{Minaev} {et~al.}(2010{\natexlab{a}}){Minaev}, {Pozanenko}, \&
  {Loznikov}}]{min10}
{Minaev}, P.~Yu., {Pozanenko}, A.~S., \& {Loznikov}, V.~M. 2010{\natexlab{a}},
  Astronomy Letters, 36, 707

\bibitem[{{Minaev} {et~al.}(2010{\natexlab{b}}){Minaev}, {Pozanenko}, \&
  {Loznikov}}]{min10b}
---. 2010{\natexlab{b}}, Astrophysical Bulletin, 65, 326

\bibitem[{{Minaev} {et~al.}(2014){Minaev}, {Pozanenko}, {Molkov}, \&
  {Grebenev}}]{min14}
{Minaev}, P.~Yu., {Pozanenko}, A.~S., {Molkov}, S.~V., \& {Grebenev}, S.~A.
  2014, Astronomy Letters, 40, 235

\bibitem[{{Moharana} \& {Piran}(2017)}]{mp17}
{Moharana}, R., \& {Piran}, T. 2017, \mnras, 472, L55

\bibitem[{{Moiseenko} {et~al.}(2006){Moiseenko}, {Bisnovatyi-Kogan}, \&
  {Ardeljan}}]{MBA06}
{Moiseenko}, S.~G., {Bisnovatyi-Kogan}, G.~S., \& {Ardeljan}, N.~V. 2006,
  \mnras, 370, 501

\bibitem[{{Nakar} \& {Sari}(2012)}]{2012ApJ...747...88N}
{Nakar}, E., \& {Sari}, R. 2012, \apj, 747, 88

\bibitem[{{Paczynski}(1986{\natexlab{a}})}]{p86}
{Paczynski}, B. 1986{\natexlab{a}}, \apjl, 308, L43

\bibitem[{{Paczynski}(1986{\natexlab{b}})}]{Paczynski86}
---. 1986{\natexlab{b}}, \apjl, 308, L43

\bibitem[{{Paczynski}(1990)}]{pacz90}
---. 1990, \apj, 363, 218

\bibitem[{{Porth} {et~al.}(2013){Porth}, {Komissarov}, \&
  {Keppens}}]{2013MNRAS.431L..48P}
{Porth}, O., {Komissarov}, S.~S., \& {Keppens}, R. 2013, \mnras, 431, L48

\bibitem[{{Porth} {et~al.}(2014){Porth}, {Komissarov}, \&
  {Keppens}}]{2014MNRAS.438..278P}
---. 2014, \mnras, 438, 278

\bibitem[{{Pozanenko} {et~al.}(2017{\natexlab{a}}){Pozanenko}, {Mazaeva},
  {Volnova}, {Minaev}, \& {Krugov}}]{chilescope-mozaic}
{Pozanenko}, A., {Mazaeva}, E., {Volnova}, A., {Minaev}, P., \& {Krugov}, M.
  2017{\natexlab{a}}, LVC GRB Coordinates Network, 21618

\bibitem[{{Pozanenko} {et~al.}(2017{\natexlab{b}}){Pozanenko}, {Volnova},
  {Mazaeva}, {Minaev}, \& {Krugov}}]{gcn21635}
{Pozanenko}, A., {Volnova}, A., {Mazaeva}, E., {Minaev}, P., \& {Krugov}, M.
  2017{\natexlab{b}}, GCN, 21635 LVC

\bibitem[{{Pozanenko} {et~al.}(2017{\natexlab{c}}){Pozanenko}, {Volnova},
  {Mazaeva}, {Minaev}, {Moskvitin}, \& {Krugov}}]{gcn21898}
{Pozanenko}, A., {Volnova}, A., {Mazaeva}, E., {Minaev}, P., {Moskvitin}, A.,
  \& {Krugov}, M. 2017{\natexlab{c}}, GCN, 21898 LVC

\bibitem[{{Pozanenko} {et~al.}(2017{\natexlab{d}}){Pozanenko}, {Volnova},
  {Minaev}, \& {Krugov}}]{gcn21644}
{Pozanenko}, A., {Volnova}, A., {Minaev}, P., \& {Krugov}, M.
  2017{\natexlab{d}}, GCN, 21644 LVC

\bibitem[{{Qin} \& {Chen}(2013)}]{qin13}
{Qin}, Y.-P., \& {Chen}, Z.-F. 2013, \mnras, 430, 163

\bibitem[{{Radice} {et~al.}(2016){Radice}, {Galeazzi}, {Lippuner}, {Roberts},
  {Ott}, \& {Rezzolla}}]{rgl16}
{Radice}, D., {Galeazzi}, F., {Lippuner}, J., {Roberts}, L.~F., {Ott}, C.~D.,
  \& {Rezzolla}, L. 2016, \mnras, 460, 3255

\bibitem[{{Rees} \& {Gunn}(1974)}]{1974MNRAS.167....1R}
{Rees}, M.~J., \& {Gunn}, J.~E. 1974, \mnras, 167, 1

\bibitem[{{Rees} \& {M{\' e}sz{\' a}ros}(2005)}]{rm05}
{Rees}, M.~J., \& {M{\' e}sz{\' a}ros}, P. 2005, The Astrophysical Journal,
  628, 847

\bibitem[{{Rezzolla} {et~al.}(2011){Rezzolla}, {Giacomazzo}, {Baiotti},
  {Granot}, {Kouveliotou}, \& {Aloy}}]{2011ApJ...732L...6R}
{Rezzolla}, L., {Giacomazzo}, B., {Baiotti}, L., {Granot}, J., {Kouveliotou},
  C., \& {Aloy}, M.~A. 2011, \apjl, 732, L6

\bibitem[{{Roberts} {et~al.}(2011){Roberts}, {Kasen}, {Lee}, \&
  {Ramirez-Ruiz}}]{rkl11}
{Roberts}, L.~F., {Kasen}, D., {Lee}, W.~H., \& {Ramirez-Ruiz}, E. 2011, \apjl,
  736, L21

\bibitem[{{Ruiz} {et~al.}(2016){Ruiz}, {Lang}, {Paschalidis}, \&
  {Shapiro}}]{rlps16}
{Ruiz}, M., {Lang}, R.~N., {Paschalidis}, V., \& {Shapiro}, S.~L. 2016, \apjl,
  824, L6

\bibitem[{{Samodurov} {et~al.}(2017){Samodurov}, {Pozanenko}, {Rodin}, \& {et
  al.}}]{samodurov17}
{Samodurov}, V., {Pozanenko}, A.~S., {Rodin}, E.~A., \& {et al.} 2017, in Data
  Analytics and Management in Data Intensive Domains. DAMDID/RCDL 2016.
  Communications in Computer and Information Science, Vol. 706, Communications in
  Computer and Information Science, ed. K.~S. {Kalinichenko}~L.

\bibitem[{{Savchenko} {et~al.}(2017{\natexlab{a}}){Savchenko}, {Ferrigno},
  {Kuulkers}, \& et~al.}]{sav17}
{Savchenko}, V., {Ferrigno}, C., {Kuulkers}, E., \& et~al. 2017{\natexlab{a}},
  \apjl, 848, L15

\bibitem[{{Savchenko} {et~al.}(2017{\natexlab{b}}){Savchenko}, {Mereghetti},
  {Ferrigno}, {Kuulkers}, \& {Bazzano}}]{gcn21507}
{Savchenko}, V., {Mereghetti}, S., {Ferrigno}, C., {Kuulkers}, E., \&
  {Bazzano}, A. 2017{\natexlab{b}}, GCN, 21507 LVC

\bibitem[{{Schlafly} \& {Finkbeiner}(2011)}]{Schlafly2011}
{Schlafly}, E.~F., \& {Finkbeiner}, D.~P. 2011, \apj, 737, 103

\bibitem[{{Shakura} \& {Sunyaev}(1973)}]{ss73}
{Shakura}, N.~I., \& {Sunyaev}, R.~A. 1973, {\aap}, 24, 337

\bibitem[{{Sikora} {et~al.}(1997){Sikora}, {Madejski}, {Moderski}, \&
  {Poutanen}}]{smmp97}
{Sikora}, M., {Madejski}, G., {Moderski}, R., \& {Poutanen}, J. 1997, \apj,
  484, 108

\bibitem[{{Smartt} {et~al.}(2017){Smartt}, {Chen}, {Jerkstrand}, \& {et
  al.}}]{2017Natur.551...75S}
{Smartt}, S.~J., {Chen}, T.-W., {Jerkstrand}, A., \& {et al.} 2017, \nat, 551,
  75

\bibitem[{{Svensson}(1982)}]{1982ApJ...258..335S}
{Svensson}, R. 1982, \apj, 258, 335

\bibitem[{{Svinkin} {et~al.}(2016){Svinkin}, {Frederiks}, {Aptekar}, \&
  et~al.}]{svi16}
{Svinkin}, D.~S., {Frederiks}, D.~D., {Aptekar}, R.~L., \& et~al. 2016, \apjs,
  224, 10

\bibitem[{{Tan} {et~al.}(2001){Tan}, {Matzner}, \&
  {McKee}}]{2001ApJ...551..946T}
{Tan}, J.~C., {Matzner}, C.~D., \& {McKee}, C.~F. 2001, \apj, 551, 946

\bibitem[{{Tanvir} \& {Levan}(2017)}]{vista-gcn}
{Tanvir}, N.~R., \& {Levan}, A.~J. 2017, GCN, 21544 LVC

\bibitem[{{Tanvir} {et~al.}(2013){Tanvir}, {Levan}, {Fruchter}, {Hjorth},
  {Hounsell}, {Wiersema}, \& {Tunnicliffe}}]{tanvir130603b}
{Tanvir}, N.~R., {Levan}, A.~J., {Fruchter}, A.~S., {Hjorth}, J., {Hounsell},
  R.~A., {Wiersema}, K., \& {Tunnicliffe}, R.~L. 2013, \nat, 500, 547

\bibitem[{{Tanvir} {et~al.}(2017){Tanvir}, {Levan}, {Gonz{\'a}lez-Fern{\'a}ndez}, \& {et~al.}}]{Tanvir2017}
{Tanvir}, N.~R., {Levan}, A.~J., {Gonz{\'a}lez-Fern{\'a}ndez}, C., \& et~al. 2017, \apjl, 848, L27

\bibitem[{{Tchekhovskoy} {et~al.}(2010){Tchekhovskoy}, {Narayan}, \&
  {McKinney}}]{tnm10}
{Tchekhovskoy}, A., {Narayan}, R., \& {McKinney}, J.~C. 2010, \na, 15, 749

\bibitem[{{Usov}(1992)}]{U92}
{Usov}, V.~V. 1992, \nat, 357, 472

\bibitem[{{Vigan{\`o}} \& {Mereghetti}(2009)}]{vig09}
{Vigan{\`o}}, D., \& {Mereghetti}, S. 2009, in The Extreme Sky: Sampling the
  Universe above 10 keV

\bibitem[{{Villar} {et~al.}(2017){Villar}, {Guillochon}, {Berger}, \&
  et~al.}]{knoptall}
{Villar}, V.~A., {Guillochon}, J., {Berger}, E., \& et~al. 2017,
  ArXiv:1710.11576

\bibitem[{{von Kienlin} {et~al.}(2003){von Kienlin}, {Beckmann}, {Rau}, \&
  et~al.}]{kienlin_03}
{von Kienlin}, A., {Beckmann}, V., {Rau}, A., \& et~al. 2003, \aap, 411, L299

\bibitem[{{von Kienlin} {et~al.}(2017){von Kienlin}, {Meegan}, \&
  {Goldstein}}]{fermi-trigger}
{von Kienlin}, A., {Meegan}, C., \& {Goldstein}, A. 2017, LVC GRB Coordinates
  Network, 21520

\bibitem[{{von Kienlin} {et~al.}(2014){von Kienlin}, {Meegan}, {Paciesas}, \&
  et~al.}]{kienlin14}
{von Kienlin}, A., {Meegan}, C.~A., {Paciesas}, W.~S., \& et~al. 2014, \apjs,
  211, 13

\bibitem[{{Wanajo} {et~al.}(2014){Wanajo}, {Sekiguchi}, {Nishimura}, {Kiuchi},
  {Kyutoku}, \& {Shibata}}]{wsn14}
{Wanajo}, S., {Sekiguchi}, Y., {Nishimura}, N., {Kiuchi}, K., {Kyutoku}, K., \&
  {Shibata}, M. 2014, \apjl, 789, L39

\bibitem[{{Weaver}(1976)}]{1976ApJS...32..233W}
{Weaver}, T.~A. 1976, \apjs, 32, 233

\bibitem[{{Wu} {et~al.}(2017){Wu}, {Tamborra}, {Just}, \& {Janka}}]{wtjj17}
{Wu}, M.-R., {Tamborra}, I., {Just}, O., \& {Janka}, H.-T. 2017, ArXiv e-prints

\bibitem[{{Yang} {et~al.}(2017){Yang}, {Valenti}, {Sand}, {Tartaglia},
  {Cappellaro}, {Reichart}, {Haislip}, \& {Kouprianov}}]{DLT-gcn}
{Yang}, S., {Valenti}, S., {Sand}, D., {Tartaglia}, L., {Cappellaro}, E.,
  {Reichart}, D., {Haislip}, J., \& {Kouprianov}, V. 2017, GCN, 21531 LVC

\bibitem[{{Zalamea} \& {Beloborodov}(2011)}]{zb11}
{Zalamea}, I., \& {Beloborodov}, A.~M. 2011, \mnras, 410, 2302

\end{thebibliography}

\appendix

\section{Optical Observations of CHILESCOPE observatory}

\subsection{The CHILESCOPE}
\label{sc:instrum}

The CHILESCOPE\footnote{\url{http://www.chilescope.com/}} is a remote controlled commercial observatory located in
the Chilean Andes (W 70.75 S 30.27) equipped with a 1 m Ritchey Chretien telescope (RC-1000) and two identical 50 cm fast Newton astrographs (Newtonian 1 ASA-500 and Newtonian 2).

Both Newtonians with $f$/3.8 are on ``$Z$'' equatorial mounts, both are equipped with 4K$\times$4K FLI PROLINE
16803 CCD cameras with Astrodon Generation 2 E-Series Luminance filter
\footnote{\url{https://www.cloudynights.com/uploads/monthly_09_2014/post-23216-0-17456300-1411330347.jpg}},
which is approximately equivalent to a clear light. The field of view is $67\times67$ arcminutes.

The 1-meter RC-1000 telescope is mounted on an alt-azimuth mount with direct drives on both axes and has a focal
ratio $f$/6.8 with a reducer. The telescope is also equipped with a 4K$\times$4K FLI PROLINE 16803 CCD camera with
Astrodon Generation 2 E-Series Luminance filter. It effectively cuts off the wavelengths $<4000$~\AA{} and $>7100$~\AA{}, thus its transmission curve corresponds to a clear light.
In the paper, we refer to this filter as Clear. The resulting field of view is 18.6$\times$18.6 arcminutes.
It also has a good thermal stabilization with the working temperature of -30$^{\circ}$C.

\subsection{Observations of the RC-1000 telescope}
\label{tab:RC1000}

The RC-1000 telescope covered the central part of the LIGO/Virgo trigger G298048
error box \citep{GW-trigger} with a typical limiting magnitude of $20^m$ at 60 s exposure
in each frame. {A total of} 48 images were obtained. The coverage of the $1\sigma$
error box of G298048 is 14.4 \%. The start time (UT) and the center {coordinates} of each
{frame are collected} in Table \ref{tab:RC}.
The optical counterpart candidate {SSS17a} was independently discovered by six teams
\citep{decam-gcn,lco-gcn,Swope-gcn,master-gcn,vista-gcn,DLT-gcn} with the first announcement at 01:05
UT Auggust 18 (12.4 hr since GW trigger and 25.6 minutes after the end of RC-1000
telescope observations). {The localization of SSS17a} is out of the coverage in the
first epoch of our observations. The covering map can be found in Figure \ref{fig:ASA}, left panel.

\defcitealias{smith2014a}{Paper~I}
\begin{table*}
\centering
 \caption{Observations of the RC-1000 telescope}
 \label{tab:RC}
 \begin{tabular}{cccc}
  \hline
	UT Date & Time Since GW Trigger (days) & Center (R.A.) & Center (Decl.) \\
  \hline
2017 Aug 17   23:44:31 & 0.46108 & 12 $^h$ 48 $^m$ 28\fs414 & -14 \degr 25 \arcmin 07\farcs037 \\ [2pt]
2017 Aug 17   23:45:53 & 0.46203 & 12 $^h$ 47 $^m$ 05\fs895 & -14 \degr 25 \arcmin 04\farcs261 \\ [2pt]
2017 Aug 17   23:47:16 & 0.46299 & 12 $^h$ 45 $^m$ 43\fs416 & -14 \degr 25 \arcmin 01\farcs827 \\ [2pt]
2017 Aug 17   23:48:38 & 0.46394 & 12 $^h$ 48 $^m$ 28\fs594 & -14 \degr 45 \arcmin 01\farcs452 \\ [2pt]
2017 Aug 17   23:50:00 & 0.46488 & 12 $^h$ 47 $^m$ 05\fs988 & -14 \degr 44 \arcmin 59\farcs153 \\ [2pt]
2017 Aug 17   23:51:23 & 0.46584 & 12 $^h$ 45 $^m$ 43\fs329 & -14 \degr 44 \arcmin 57\farcs545 \\ [2pt]
2017 Aug 17   23:52:45 & 0.46679 & 12 $^h$ 48 $^m$ 28\fs701 & -15 \degr 04 \arcmin 58\farcs286 \\ [2pt]
2017 Aug 17   23:54:07 & 0.46774 & 12 $^h$ 47 $^m$ 05\fs854 & -15 \degr 04 \arcmin 58\farcs530 \\ [2pt]
2017 Aug 17   23:55:30 & 0.46870 & 12 $^h$ 45 $^m$ 43\fs005 & -15 \degr 04 \arcmin 59\farcs031 \\ [2pt]
2017 Aug 17   23:58:01 & 0.47045 & 12 $^h$ 51 $^m$ 28\fs799 & -15 \degr 25 \arcmin 03\farcs691 \\ [2pt]
2017 Aug 17   23:59:23 & 0.47140 & 12 $^h$ 50 $^m$ 05\fs874 & -15 \degr 25 \arcmin 02\farcs099 \\ [2pt]
2017 Aug 18   00:00:46 & 0.47236 & 12 $^h$ 48 $^m$ 42\fs901 & -15 \degr 25 \arcmin 00\farcs518 \\ [2pt]
2017 Aug 18   00:02:17 & 0.47341 & 12 $^h$ 51 $^m$ 28\fs942 & -15 \degr 45 \arcmin 00\farcs766 \\ [2pt]
2017 Aug 18   00:03:49 & 0.47448 & 12 $^h$ 50 $^m$ 05\fs842 & -15 \degr 44 \arcmin 59\farcs895 \\ [2pt]
2017 Aug 18   00:05:22 & 0.47556 & 12 $^h$ 48 $^m$ 42\fs695 & -15 \degr 44 \arcmin 56\farcs860 \\ [2pt]
2017 Aug 18   00:06:53 & 0.47661 & 12 $^h$ 51 $^m$ 28\fs907 & -16 \degr 04 \arcmin 59\farcs101 \\ [2pt]
2017 Aug 18   00:08:15 & 0.47756 & 12 $^h$ 50 $^m$ 05\fs657 & -16 \degr 04 \arcmin 57\farcs618 \\ [2pt]
2017 Aug 18   00:09:38 & 0.47852 & 12 $^h$ 48 $^m$ 42\fs360 & -16 \degr 04 \arcmin 57\farcs688 \\ [2pt]
2017 Aug 18   00:13:16 & 0.48104 & 12 $^h$ 51 $^m$ 28\fs630 & -15 \degr 25 \arcmin 05\farcs288 \\ [2pt]
2017 Aug 18   00:14:38 & 0.48199 & 12 $^h$ 50 $^m$ 05\fs706 & -15 \degr 25 \arcmin 02\farcs368 \\ [2pt]
2017 Aug 18   00:17:15 & 0.48381 & 12 $^h$ 54 $^m$ 29\fs036 & -16 \degr 25 \arcmin 02\farcs125 \\ [2pt]
2017 Aug 18   00:18:50 & 0.48491 & 12 $^h$ 54 $^m$ 29\fs003 & -16 \degr 25 \arcmin 03\farcs603 \\ [2pt]
2017 Aug 18   00:20:21 & 0.48596 & 12 $^h$ 53 $^m$ 05\fs630 & -16 \degr 25 \arcmin 02\farcs356 \\ [2pt]
2017 Aug 18   00:21:52 & 0.48701 & 12 $^h$ 51 $^m$ 42\fs259 & -16 \degr 25 \arcmin 00\farcs875 \\ [2pt]
2017 Aug 18   00:23:22 & 0.48806 & 12 $^h$ 54 $^m$ 29\fs055 & -16 \degr 45 \arcmin 00\farcs484 \\ [2pt]
2017 Aug 18   00:24:53 & 0.48911 & 12 $^h$ 53 $^m$ 05\fs560 & -16 \degr 44 \arcmin 58\farcs705 \\ [2pt]
2017 Aug 18   00:26:26 & 0.49019 & 12 $^h$ 51 $^m$ 41\fs997 & -16 \degr 44 \arcmin 57\farcs424 \\ [2pt]
2017 Aug 18   00:27:57 & 0.49124 & 12 $^h$ 54 $^m$ 29\fs024 & -17 \degr 04 \arcmin 59\farcs058 \\ [2pt]
2017 Aug 18   00:29:20 & 0.49220 & 12 $^h$ 53 $^m$ 05\fs393 & -17 \degr 04 \arcmin 59\farcs049 \\ [2pt]
2017 Aug 18   00:30:42 & 0.49315 & 12 $^h$ 51 $^m$ 41\fs709 & -17 \degr 04 \arcmin 58\farcs549 \\ [2pt]
2017 Aug 18   00:34:03 & 0.49547 & 12 $^h$ 57 $^m$ 29\fs158 & -17 \degr 25 \arcmin 03\farcs586 \\ [2pt]
2017 Aug 18   00:35:33 & 0.49652 & 12 $^h$ 56 $^m$ 05\fs408 & -17 \degr 25 \arcmin 02\farcs790 \\ [2pt]
2017 Aug 18   00:37:04 & 0.49757 & 12 $^h$ 54 $^m$ 41\fs653 & -17 \degr 24 \arcmin 59\farcs856 \\ [2pt]
2017 Aug 18   00:38:26 & 0.49852 & 12 $^h$ 57 $^m$ 29\fs218 & -17 \degr 45 \arcmin 01\farcs451 \\ [2pt]
2017 Aug 18   00:39:49 & 0.49948 & 12 $^h$ 56 $^m$ 05\fs305 & -17 \degr 44 \arcmin 59\farcs001 \\ [2pt]
2017 Aug 18   00:41:19 & 0.50052 & 12 $^h$ 54 $^m$ 41\fs310 & -17 \degr 44 \arcmin 57\farcs055 \\ [2pt]
2017 Aug 18   00:42:50 & 0.50157 & 12 $^h$ 57 $^m$ 29\fs230 & -18 \degr 04 \arcmin 59\farcs511 \\ [2pt]
2017 Aug 18   00:44:13 & 0.50253 & 12 $^h$ 56 $^m$ 05\fs075 & -18 \degr 04 \arcmin 58\farcs201 \\ [2pt]
2017 Aug 18   00:45:35 & 0.50348 & 12 $^h$ 54 $^m$ 40\fs882 & -18 \degr 04 \arcmin 57\farcs991 \\ [2pt]
2017 Aug 18   00:48:13 & 0.50531 & 13 $^h$ 00 $^m$ 29\fs457 & -18 \degr 25 \arcmin 01\farcs845 \\ [2pt]
2017 Aug 18   00:49:43 & 0.50635 & 12 $^h$ 59 $^m$ 05\fs148 & -18 \degr 25 \arcmin 01\farcs236 \\ [2pt]
2017 Aug 18   00:51:06 & 0.50731 & 12 $^h$ 57 $^m$ 40\fs809 & -18 \degr 24 \arcmin 59\farcs051 \\ [2pt]
2017 Aug 18   00:52:28 & 0.50826 & 13 $^h$ 00 $^m$ 29\fs458 & -18 \degr 44 \arcmin 59\farcs417 \\ [2pt]
2017 Aug 18   00:53:50 & 0.50921 & 12 $^h$ 59 $^m$ 04\fs993 & -18 \degr 44 \arcmin 57\farcs667 \\ [2pt]
2017 Aug 18   00:55:23 & 0.51029 & 12 $^h$ 57 $^m$ 40\fs424 & -18 \degr 44 \arcmin 56\farcs776 \\ [2pt]
2017 Aug 18   00:56:54 & 0.51134 & 13 $^h$ 00 $^m$ 29\fs415 & -19 \degr 04 \arcmin 57\farcs375 \\ [2pt]
2017 Aug 18   00:58:25 & 0.51240 & 12 $^h$ 59 $^m$ 04\fs778 & -19 \degr 04 \arcmin 57\farcs208 \\ [2pt]
2017 Aug 18   00:59:57 & 0.51346 & 12 $^h$ 57 $^m$ 40\fs097 & -19 \degr 04 \arcmin 55\farcs261 \\ [2pt]
  \hline
 \end{tabular}
\end{table*}

\subsection{Observations the of ASA-500 telescope}
\label{sec:ASA500}

\begin{figure}
\includegraphics[width=1.0\linewidth,angle=-0]{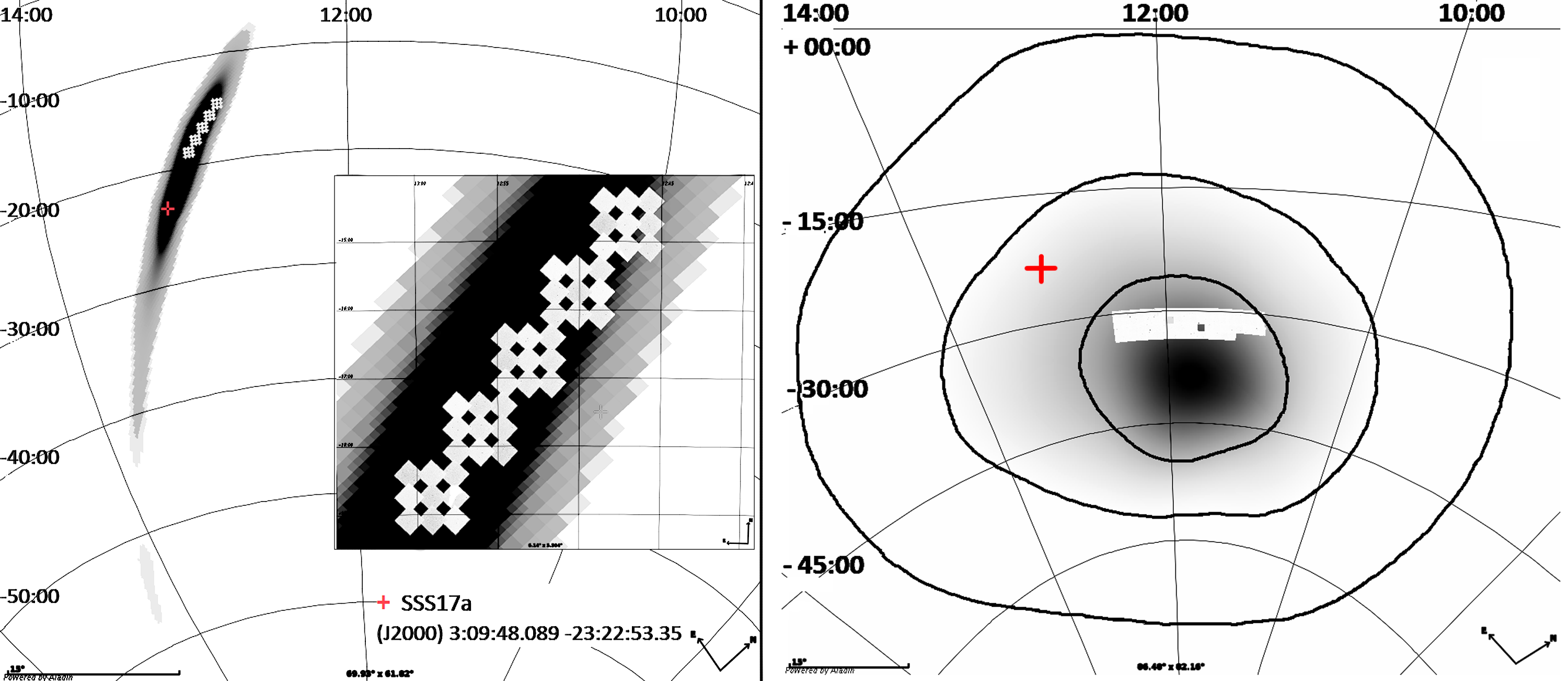}
\caption{Mosaic observations of the RC-1000 telescope (left panel) and the ASA-500 telescope (right panel). The figure shows the LIGO/Virgo error box with partial coverage by the RC-1000 telescope observations (left panel). Observations of the RC-1000 telescope began 11.1 hr after the GW trigger. Coverage of ASA-500 telescope observations of the GBM/\textit{Fermi} localization area of GRB~170817A  is shown in the right panel. Localization area at 1, 2, and 3$\sigma$ are shown by thick curves. Observations of the ASA-500 telescope began 10.6 hr after the GW trigger. The red cross is the position
of the optical counterpart (R.A.(J2000)=03:09:48.089, Decl.(J2000)=-23:22:53.35).}
\label{fig:ASA}
\end{figure}

We also covered the northern part of the GBM/\textit{Fermi} localization $1\sigma$ containment \citep[statistical only, ][]{gcn21506} with several frames taken by the ASA-500 telescope in the Clear filter. The total coverage of the GBM $1\sigma$ localization area is about 16~\% and contains 76 frames. The typical limiting magnitude {of a single frame is} $17\fm5$ at 60 s exposure. Date, start time of observations (UT), and center coordinates of each frame are collected in Tables~\ref{tab:ASA} and \ref{tab:continued}. The covering map can be found in Figure~\ref{fig:ASA}, right panel.

\defcitealias{smith2014b}{Paper~I}
\begin{table*}
\centering
 \caption{Observations of ASA-500 telescope}
 \label{tab:ASA}
 \begin{tabular}{cccc}
  \hline
	UT Date & Time Since GW Trigger (days) & Center (R.A.) & Center (Decl.) \\
  \hline
2017 Aug 17   23:17:16 & 0.44215 & 12 $^h$ 30 $^m$ 15\fs598 & -30 \degr 12 \arcmin 47\farcs661 \\ [2pt]
2017 Aug 17   23:18:42 & 0.44315 & 12 $^h$ 25 $^m$ 37\fs552 & -30 \degr 12 \arcmin 46\farcs982 \\ [2pt]
2017 Aug 17   23:20:08 & 0.44414 & 12 $^h$ 20 $^m$ 59\fs825 & -30 \degr 12 \arcmin 51\farcs272 \\ [2pt]
2017 Aug 17   23:21:34 & 0.44514 & 12 $^h$ 16 $^m$ 22\fs006 & -30 \degr 12 \arcmin 54\farcs690 \\ [2pt]
2017 Aug 17   23:23:00 & 0.44613 & 12 $^h$ 11 $^m$ 44\fs220 & -30 \degr 12 \arcmin 56\farcs952 \\ [2pt]
2017 Aug 17   23:24:26 & 0.44713 & 12 $^h$ 07 $^m$ 06\fs472 & -30 \degr 12 \arcmin 59\farcs341 \\ [2pt]
2017 Aug 17   23:25:51 & 0.44811 & 12 $^h$ 02 $^m$ 28\fs537 & -30 \degr 13 \arcmin 04\farcs766 \\ [2pt]
2017 Aug 17   23:27:08 & 0.44900 & 11 $^h$ 57 $^m$ 50\fs769 & -30 \degr 13 \arcmin 06\farcs317 \\ [2pt]
2017 Aug 17   23:28:26 & 0.44991 & 11 $^h$ 53 $^m$ 12\fs785 & -30 \degr 13 \arcmin 12\farcs156 \\ [2pt]
2017 Aug 17   23:29:51 & 0.45089 & 11 $^h$ 48 $^m$ 34\fs883 & -30 \degr 13 \arcmin 16\farcs196 \\ [2pt]
2017 Aug 17   23:31:09 & 0.45179 & 11 $^h$ 43 $^m$ 56\fs993 & -30 \degr 13 \arcmin 19\farcs841 \\ [2pt]
2017 Aug 17   23:32:26 & 0.45269 & 11 $^h$ 38 $^m$ 46\fs482 & -31 \degr 13 \arcmin 44\farcs353 \\ [2pt]
2017 Aug 17   23:33:51 & 0.45367 & 11 $^h$ 34 $^m$ 41\fs269 & -30 \degr 13 \arcmin 27\farcs977 \\ [2pt]
2017 Aug 17   23:35:09 & 0.45457 & 11 $^h$ 30 $^m$ 03\fs364 & -30 \degr 13 \arcmin 34\farcs495 \\ [2pt]
2017 Aug 17   23:36:26 & 0.45546 & 11 $^h$ 25 $^m$ 25\fs436 & -30 \degr 13 \arcmin 37\farcs753 \\ [2pt]
2017 Aug 17   23:37:51 & 0.45645 & 11 $^h$ 20 $^m$ 47\fs749 & -30 \degr 13 \arcmin 41\farcs649 \\ [2pt]
2017 Aug 17   23:39:16 & 0.45743 & 11 $^h$ 16 $^m$ 09\fs869 & -30 \degr 13 \arcmin 48\farcs310 \\ [2pt]
2017 Aug 17   23:40:34 & 0.45833 & 11 $^h$ 11 $^m$ 31\fs887 & -30 \degr 13 \arcmin 52\farcs435 \\ [2pt]
2017 Aug 17   23:41:51 & 0.45922 & 11 $^h$ 06 $^m$ 54\fs122 & -30 \degr 13 \arcmin 57\farcs672 \\ [2pt]
2017 Aug 17   23:43:16 & 0.46021 & 11 $^h$ 02 $^m$ 16\fs046 & -30 \degr 14 \arcmin 03\farcs015 \\ [2pt]
2017 Aug 17   23:44:34 & 0.46111 & 12 $^h$ 30 $^m$ 15\fs005 & -30 \degr 13 \arcmin 04\farcs766 \\ [2pt]
2017 Aug 17   23:45:51 & 0.46200 & 12 $^h$ 25 $^m$ 33\fs996 & -31 \degr 13 \arcmin 07\farcs322 \\ [2pt]
2017 Aug 17   23:47:17 & 0.46300 & 12 $^h$ 20 $^m$ 53\fs256 & -31 \degr 13 \arcmin 10\farcs210 \\ [2pt]
2017 Aug 17   23:48:43 & 0.46399 & 12 $^h$ 16 $^m$ 12\fs320 & -31 \degr 13 \arcmin 11\farcs223 \\ [2pt]
2017 Aug 17   23:50:09 & 0.46499 & 12 $^h$ 11 $^m$ 31\fs649 & -31 \degr 13 \arcmin 15\farcs275 \\ [2pt]
2017 Aug 17   23:51:36 & 0.46600 & 12 $^h$ 06 $^m$ 50\fs885 & -31 \degr 13 \arcmin 18\farcs115 \\ [2pt]
2017 Aug 17   23:53:02 & 0.46699 & 12 $^h$ 02 $^m$ 10\fs096 & -31 \degr 13 \arcmin 24\farcs372 \\ [2pt]
2017 Aug 17   23:54:28 & 0.46799 & 11 $^h$ 57 $^m$ 29\fs387 & -31 \degr 13 \arcmin 26\farcs510 \\ [2pt]
2017 Aug 17   23:55:45 & 0.46888 & 11 $^h$ 52 $^m$ 48\fs745 & -31 \degr 13 \arcmin 31\farcs809 \\ [2pt]
2017 Aug 17   23:57:02 & 0.46977 & 11 $^h$ 48 $^m$ 07\fs899 & -31 \degr 13 \arcmin 33\farcs279 \\ [2pt]
2017 Aug 17   23:58:28 & 0.47076 & 11 $^h$ 43 $^m$ 27\fs248 & -31 \degr 13 \arcmin 38\farcs379 \\ [2pt]
2017 Aug 17   23:59:53 & 0.47175 & 11 $^h$ 38 $^m$ 12\fs086 & -32 \degr 14 \arcmin 00\farcs137 \\ [2pt]
2017 Aug 18   00:01:18 & 0.47273 & 11 $^h$ 34 $^m$ 05\fs570 & -31 \degr 13 \arcmin 47\farcs746 \\ [2pt]
2017 Aug 18   00:02:35 & 0.47362 & 11 $^h$ 29 $^m$ 24\fs922 & -31 \degr 13 \arcmin 51\farcs547 \\ [2pt]
2017 Aug 18   00:04:00 & 0.47461 & 11 $^h$ 24 $^m$ 44\fs156 & -31 \degr 13 \arcmin 57\farcs384 \\ [2pt]
2017 Aug 18   00:05:18 & 0.47551 & 11 $^h$ 20 $^m$ 03\fs402 & -31 \degr 14 \arcmin 02\farcs252 \\ [2pt]
2017 Aug 18   00:06:35 & 0.47640 & 11 $^h$ 15 $^m$ 22\fs440 & -31 \degr 14 \arcmin 06\farcs208 \\ [2pt]
2017 Aug 18   00:08:00 & 0.47738 & 11 $^h$ 10 $^m$ 41\fs821 & -31 \degr 14 \arcmin 09\farcs616 \\ [2pt]
2017 Aug 18   00:09:18 & 0.47829 & 11 $^h$ 06 $^m$ 01\fs019 & -31 \degr 14 \arcmin 16\farcs647 \\ [2pt]
2017 Aug 18   00:10:35 & 0.47918 & 11 $^h$ 01 $^m$ 20\fs029 & -31 \degr 14 \arcmin 19\farcs995 \\ [2pt]
2017 Aug 18   00:12:05 & 0.48022 & 12 $^h$ 30 $^m$ 13\fs856 & -32 \degr 13 \arcmin 21\farcs158 \\ [2pt]
2017 Aug 18   00:13:23 & 0.48112 & 12 $^h$ 25 $^m$ 30\fs032 & -32 \degr 13 \arcmin 23\farcs827 \\ [2pt]
2017 Aug 18   00:14:40 & 0.48201 & 12 $^h$ 20 $^m$ 46\fs426 & -32 \degr 13 \arcmin 25\farcs644 \\ [2pt]
2017 Aug 18   00:16:06 & 0.48301 & 12 $^h$ 16 $^m$ 02\fs484 & -32 \degr 13 \arcmin 29\farcs700 \\ [2pt]
2017 Aug 18   00:17:23 & 0.48390 & 12 $^h$ 11 $^m$ 18\fs718 & -32 \degr 13 \arcmin 32\farcs247 \\ [2pt]
2017 Aug 18   00:18:40 & 0.48479 & 12 $^h$ 06 $^m$ 35\fs049 & -32 \degr 13 \arcmin 34\farcs391 \\ [2pt]
2017 Aug 18   00:19:58 & 0.48569 & 12 $^h$ 01 $^m$ 51\fs128 & -32 \degr 13 \arcmin 39\farcs111 \\ [2pt]
2017 Aug 18   00:21:15 & 0.48659 & 11 $^h$ 57 $^m$ 07\fs348 & -32 \degr 13 \arcmin 43\farcs128 \\ [2pt]
2017 Aug 18   00:22:32 & 0.48748 & 11 $^h$ 52 $^m$ 23\fs580 & -32 \degr 13 \arcmin 47\farcs111 \\ [2pt]
2017 Aug 18   00:23:57 & 0.48846 & 11 $^h$ 47 $^m$ 39\fs710 & -32 \degr 13 \arcmin 50\farcs407 \\ [2pt]
2017 Aug 18   00:25:15 & 0.48936 & 11 $^h$ 42 $^m$ 56\fs111 & -32 \degr 13 \arcmin 54\farcs176 \\ [2pt]
  \hline
 \end{tabular}
\end{table*}

\begin{table*}
\centering
 \caption{Observations of ASA-500 telescope, continued}
 \label{tab:continued}
 \begin{tabular}{cccc}
  \hline
	UT Date & Time Since GW Trigger (days) & Center (R.A.) & Center (Decl.) \\
  \hline
2017 Aug 18   00:27:57 & 0.49124 & 11 $^h$ 33 $^m$ 28\fs248 & -32 \degr 14 \arcmin 05\farcs578 \\ [2pt]
2017 Aug 18   00:29:22 & 0.49222 & 11 $^h$ 28 $^m$ 44\fs497 & -32 \degr 14 \arcmin 08\farcs661 \\ [2pt]
2017 Aug 18   00:30:39 & 0.49311 & 11 $^h$ 24 $^m$ 00\fs649 & -32 \degr 14 \arcmin 14\farcs154 \\ [2pt]
2017 Aug 18   00:31:57 & 0.49402 & 11 $^h$ 19 $^m$ 16\fs820 & -32 \degr 14 \arcmin 20\farcs406 \\ [2pt]
2017 Aug 18   00:33:14 & 0.49491 & 11 $^h$ 14 $^m$ 32\fs974 & -32 \degr 14 \arcmin 23\farcs645 \\ [2pt]
2017 Aug 18   00:34:31 & 0.49580 & 11 $^h$ 09 $^m$ 49\fs015 & -32 \degr 14 \arcmin 27\farcs265 \\ [2pt]
2017 Aug 18   00:35:58 & 0.49681 & 11 $^h$ 05 $^m$ 05\fs077 & -32 \degr 14 \arcmin 30\farcs945 \\ [2pt]
2017 Aug 18   00:37:24 & 0.49780 & 11 $^h$ 00 $^m$ 21\fs126 & -32 \degr 14 \arcmin 35\farcs378 \\ [2pt]
2017 Aug 18   00:38:54 & 0.49884 & 12 $^h$ 30 $^m$ 13\fs123 & -33 \degr 13 \arcmin 39\farcs301 \\ [2pt]
2017 Aug 18   00:40:21 & 0.49985 & 12 $^h$ 25 $^m$ 26\fs171 & -33 \degr 13 \arcmin 37\farcs855 \\ [2pt]
2017 Aug 18   00:41:38 & 0.50074 & 12 $^h$ 20 $^m$ 38\fs975 & -33 \degr 13 \arcmin 44\farcs215 \\ [2pt]
2017 Aug 18   00:42:55 & 0.50163 & 12 $^h$ 15 $^m$ 52\fs081 & -33 \degr 13 \arcmin 45\farcs463 \\ [2pt]
2017 Aug 18   00:44:12 & 0.50252 & 12 $^h$ 11 $^m$ 05\fs080 & -33 \degr 13 \arcmin 48\farcs273 \\ [2pt]
2017 Aug 18   00:45:38 & 0.50352 & 12 $^h$ 06 $^m$ 18\fs116 & -33 \degr 13 \arcmin 51\farcs389 \\ [2pt]
2017 Aug 18   00:47:04 & 0.50451 & 12 $^h$ 01 $^m$ 31\fs219 & -33 \degr 13 \arcmin 57\farcs070 \\ [2pt]
2017 Aug 18   00:48:30 & 0.50551 & 11 $^h$ 56 $^m$ 44\fs050 & -33 \degr 14 \arcmin 00\farcs839 \\ [2pt]
2017 Aug 18   00:49:56 & 0.50650 & 11 $^h$ 51 $^m$ 56\fs967 & -33 \degr 14 \arcmin 04\farcs169 \\ [2pt]
2017 Aug 18   00:51:21 & 0.50749 & 11 $^h$ 47 $^m$ 10\fs057 & -33 \degr 14 \arcmin 07\farcs798 \\ [2pt]
2017 Aug 18   00:52:47 & 0.50848 & 11 $^h$ 39 $^m$ 19\fs229 & -30 \degr 13 \arcmin 24\farcs617 \\ [2pt]
2017 Aug 18   00:52:47 & 0.50848 & 11 $^h$ 42 $^m$ 22\fs994 & -33 \degr 14 \arcmin 11\farcs486 \\ [2pt]
2017 Aug 18   00:54:13 & 0.50948 & 11 $^h$ 37 $^m$ 35\fs838 & -33 \degr 14 \arcmin 16\farcs286 \\ [2pt]
2017 Aug 18   00:55:39 & 0.51047 & 11 $^h$ 32 $^m$ 49\fs002 & -33 \degr 14 \arcmin 20\farcs323 \\ [2pt]
2017 Aug 18   00:57:05 & 0.51147 & 11 $^h$ 28 $^m$ 01\fs904 & -33 \degr 14 \arcmin 23\farcs832 \\ [2pt]
2017 Aug 18   00:58:31 & 0.51247 & 11 $^h$ 23 $^m$ 14\fs613 & -33 \degr 14 \arcmin 28\farcs525 \\ [2pt]
2017 Aug 18   00:59:48 & 0.51336 & 11 $^h$ 18 $^m$ 27\fs436 & -33 \degr 14 \arcmin 32\farcs516 \\ [2pt]
  \hline
 \end{tabular}
\end{table*}

\subsection{SSS17a observations}
\label{sec:SSS17a}
	
\begin{figure}
\includegraphics[width=1.0\linewidth,angle=-0]{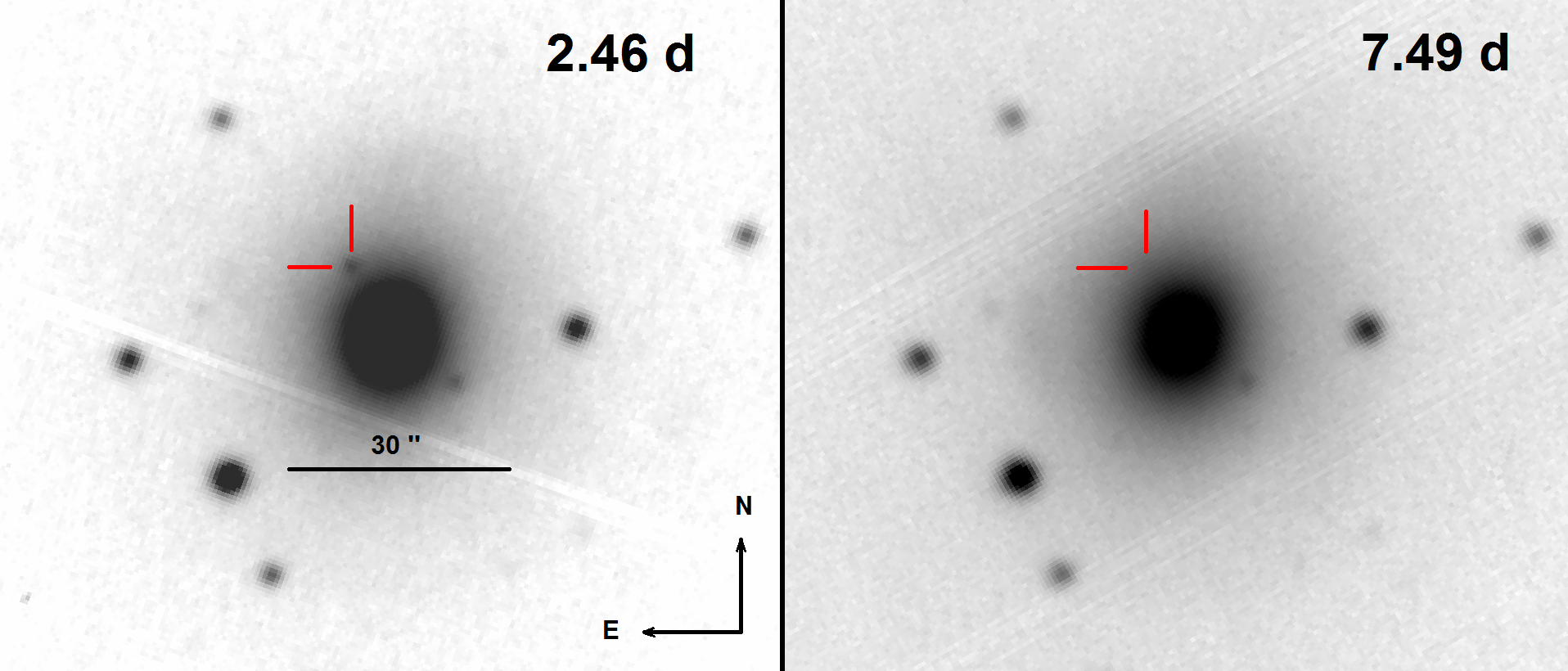}
\caption{Imaging of the SSS17a / AT2017gfo in the outskirts of the NGC 4993 galaxy with Chilescope/RC-1000 in two different epochs. Exposures of both frames are $10\times180$ s in the Clear filter.}
\label{fig:OTMAP}
\end{figure}

We started to observe the optical counterpart of the GW170817 \citep{GW-trigger} labeled SSS17a \citep{Swope-gcn} on 2017 August 19
at 23:30:33 UT taking several 180 s exposures in the Clear filter with the RC-1000 telescope of the CHILESCOPE observatory
 \citep{gcn21635}. We clearly detected the source strongly contaminated by the host
galaxy NGC~4993 background. We continued our observations on 2017 August 20, 21, and 24, taking several 180 s
exposures in each epoch \citep{gcn21644,gcn21898}. The weather conditions were satisfactory with a median
seeing of 2 arcsec. The seeing was not good for the Chilean sky because the source was observed in the end of
twilight when the atmosphere was not stable. The source was clearly detected in the stacked frames on August 20
and 21   but faded on August 24 below the detection limit over a host galaxy background. The optical transient SSS17a is shown in  Figure~\ref{fig:OTMAP}.  The log of our observations is listed in the Table~\ref{log}.

\begin{table*}
\centering
 \begin{minipage}{140mm}
  \caption{Log of the Optical Observations and Photometry of the GW170817. }
  \begin{tabular}{llllccl}
\hline
Date & UT Start & Exposure, s & Elapsed Time, days & $R$ mag. & Error \\
\hline
2017 Aug 19 & 23:30:33 & 10$\times$180 & 2.46157  &   19.12  & 0.06 \\
2017 Aug 20 & 23:21:09 & 13$\times$180 & 3.47840  &   20.04  & 0.08 \\
2017 Aug 21 & 23:32:09 & 22$\times$180 & 4.49409  &   20.14  & 0.12 \\
2017 Aug 24 & 23:53:39 & 20$\times$180 & 7.49278  &  $>21.0$ & --   \\
\hline
 \end{tabular}
  \label{log}
  \newline \textbf{Note.} Observations in Clear filter were calibrated against USNO-B1.0  Stars, R2 magnitudes.
  The magnitudes in the table are not corrected for the Galactic extinction due to the reddening of E(B-V) = 0.109 \cite{Schlafly2011} in the direction of the burst.
\end{minipage}
\end{table*}
	
\subsection{Data Reduction and Photometry}
\label{sec:photometry}
	
	All primary reduction of CCD images (bias and dark subtraction, flat-fielding) was performed using the CCDPROC
task of the IRAF software package\footnote{\url{http://iraf.noao.edu/}}. The flux of the source is strongly
affected by the host galaxy contribution, so the strategy of direct image subtraction was decided. Since the
source was not detected in the stacked frame of the last observational epoch (August 24), we chose this epoch as
a template for the subtraction of the host galaxy. To align images of different observational epochs, we used the
package ALIGN/IMAGE of ESO-MIDAS software\footnote{\url{http://www.eso.org/sci/software/esomidas/}} and the
\texttt{geomap} task of IRAF. The average image background was subtracted from all frames using median filter,
and flux normalization for the image subtraction was performed using the MAGNITUDE/CIRCLE task of ESO-MIDAS.
	
	The photometry of the source after the host subtraction was made with the PSF method using the DAOPHOT package
from IRAF software. A rectangular area containing the host galaxy and the source SSS17a was replaced
in the stacked background-subtracted frames with the host subtracted flux-normalized sub-images, made for
each observational epoch. The reference PSF stars for each specific epoch were taken from the area outside
of the host subtraction region.
	
The resulting instrumental photometrical magnitudes were calibrated with the USNO-B1.0 catalog R2 magnitudes.
After the study of stellar non-saturated objects in the field, we chose four reference stars.
The information about them is listed in the Table~\ref{refstars}. The results of the photometry are
presented in Table~\ref{log}.

\subsection{Optical light curve}
\label{sec:lightcurve}

\begin{figure}
\includegraphics[width=175mm,angle=-0]{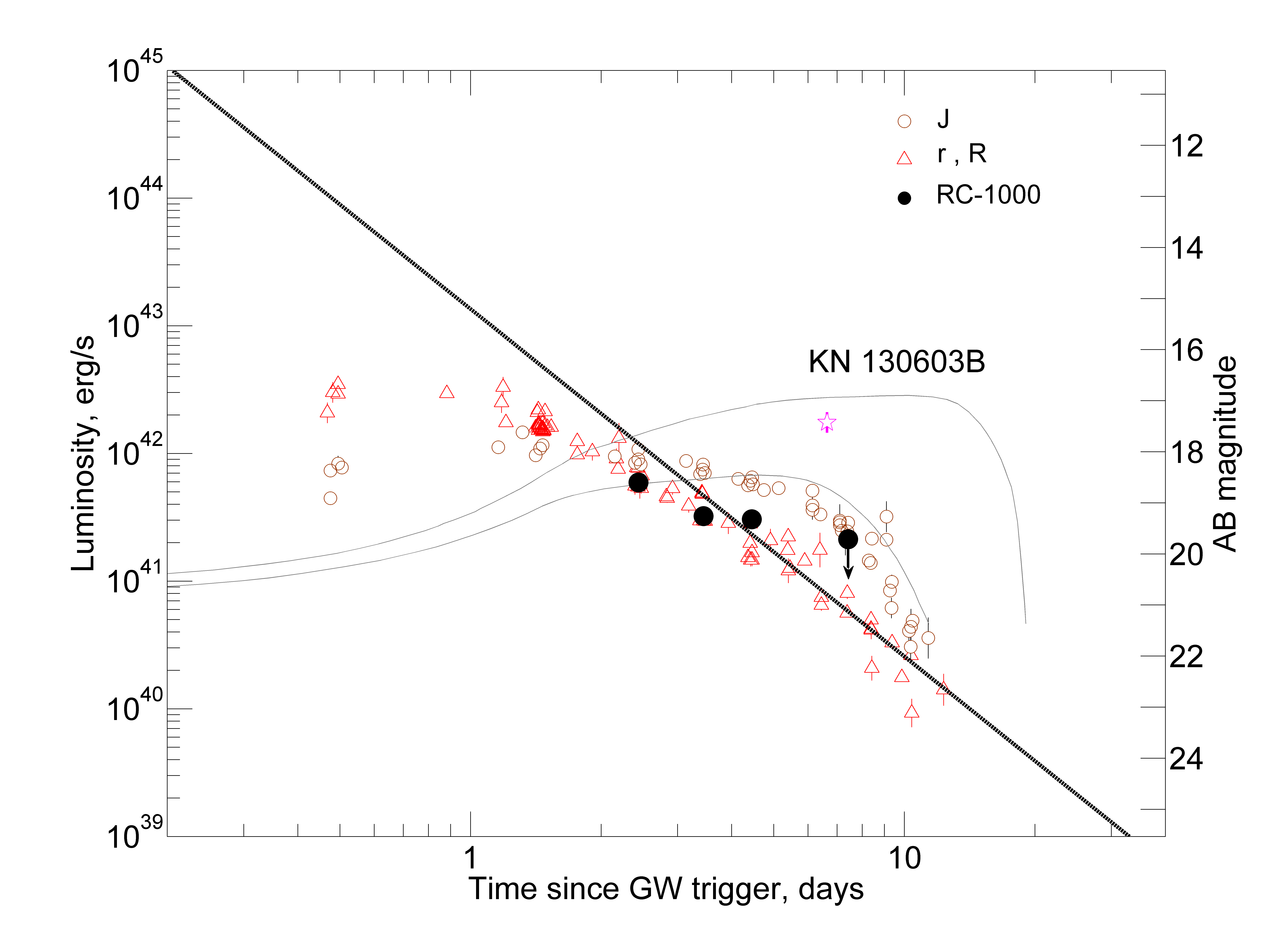}
\caption{Light curve of the OT in units of stellar magnitudes (right Y-axis) and total luminosity $L_{\rm iso}$ (left Y-axis). Bold black circles represent the observations made by the CHILESCOPE/RC-1000 telescope. Red open triangles and brown open circles show $r$-, $R$-, and $J$-band data, respectively, taken from \citet{knoptall,Tanvir2017}. The optical afterglow approximation of GRB~130603B rescaled to the rest frame is shown by the dashed line. The luminosity of a kilonova associated with GRB~130603B is also shown (purple star).
The luminosity is calculated from nIR data obtained in the F160W
filter of HST \citep{tanvir130603b}. Thin black curves represent the models from \citet{tanvir130603b}, corresponding to ejected masses of $10^{-2} M_{\odot}$ [lower] and $10^{-1} M_{\odot}$ [upper] respectively. }
\label{fig:kilonova+AG}
\end{figure}

The light curve of the optical transient, including published so far photometry collected in \cite{knoptall,Tanvir2017} in $R, r$ and in $J$ filters is presented the Figure \ref{fig:kilonova+AG}. Bold black circles represent the data obtained by RC-1000 telescope in the Clear filter. The data were shifted by 0.25 mag to compensate for the difference between the photometric bands. The shift was calculated using $r$ or $R$ data obtained simultaneously with ours. In the same figure, we plot the
nIR afterglow of GRB~130603B (rescaled in both frequency and time) to the rest frame. We used parameters of the afterglow approximation (Flux $\sim t^{-2.72}$) after jet-break at about 0.4 days \citep{tanvir130603b}. The absolute calibration in the rest frame was calculated from the broadband SED of GRB~130603B obtained at $\sim 0.6$ days post-burst (in the observer frame) in filters $grizJK$ \citep[Table 2 in ][]{tanvir130603b}. The plotted afterglow roughly corresponds to the $J$-filter in the rest frame. For correct calculation of luminosity light curves of GRB~170817A and GRB~130603B we used Galactic extinction in the direction of the bursts, $E(B - V) = 0.109$ and $E(B - V) = 0.02$, correspondingly \citep{Schlafly2011}. A host galaxy extinction was not taken into account. It is evident from Figure \ref{fig:kilonova+AG} that the afterglow luminosity of GRB~170817A is more than 130 times fainter than that of  GRB~130603B in the $J$-filter.

\begin{table*}
\centering
  \caption{Reference Stars Used for the Photometrical Reduction.}
  \begin{tabular}{lllcc}
	\hline
	USNO-B1.0 ID & R.A. & Decl. & USON-B1.0 R2 mag. & PanSTARRS r mag. (err) \\
	\hline
	0665-0279047 & 13:09:34.72 & -23:24:45.9 & 17.26 & 17.474 (0.004) \\
	0666-0290876 & 13:09:49.74 & -23:20:34.6 & 16.51 & 16.812 (0.006) \\
	0666-0290916 & 13:10:04.58 & -23:20:47.8 & 17.23 & 17.466 (0.009) \\
	0665-0279123 & 13:09:54.37 & -23:25:34.3 & 16.89 & 17.224 (0.014) \\
	\hline
  \end{tabular}
  \label{refstars}
\end{table*}

\section{BSA: radio observations at 110~MHz}
\label{sc:BSA}

One of the most sensitive radio telescopes at the frequency of 110 MHz
is the Big Scanning Antenna (BSA, Puschino, Russia). BSA is a radio telescope of meridian type. The BSA observation is a
continuous survey in multibeam mode in the frequency range of 109.0--111.5 MHz using 96 beams covering a field
of view from $-8$ and up to $+42^{\circ}$ in decl.  with a time resolution of 12.5~ms \citep{samodurov17}.

The form of the BSA single beam diagram is described by the function
\be
(\alpha,\delta) = \left(\frac{\sin x}{x}\right)^{2}\cdot \left(\frac{\sin y}{y}\right)^{2},
\label{eq:radio_diagram}
\ee
where $x =(\pi D_{1}/\lambda)\cdot (\alpha - \alpha_{0})$, $y = (\pi   D_{2}\cos(Z)/\lambda)\cdot (\delta - \delta_{0})$, $\alpha$ and $\delta$ are  equatorial coordinates, $(\alpha_{0}, \delta_{0})$ are the coordinates determining the position of the maximum of the diagram directionality of the radio telescope, $D_{1}$ = 384 m and $D_{2}$ = 187 m are the dimensions of BSA in the direction from north to south, and from  east to west, respectively, $\lambda$ = 2.72 m is a wavelength, $Z$ is the angle between direction to object and antenna normal. It can be found from  (\ref{eq:radio_diagram})  that the size of the primary beam is about 50 arcmin in R.A. (or nearly 5 minutes) and about 30 arcmin  in  decl.

The OT position at the time of the GW170817 trigger was 11$^{\circ}$.5 above
horizon at the BSA location, but outside the BSA multibeam diagram, that is, 15$^{\circ}$ lower in decl. and 4$^{\circ}$ 
(or 14 minutes in R.A.) east from the BSA pointing direction.  BSA can detect radio transients around twentieth side lobes of southern beams if the transient would be sufficiently bright. It can be found from (\ref{eq:radio_diagram}) that the twentieth side lobe has an efficiency of about $3\cdot 10^{-4}$. However, (\ref{eq:radio_diagram}) is the formula for ideal antenna.
{Really,} bright radio sources
such as Sun or Cygnus A (3C 405) are indeed registered in side lobes of BSA at a distance of $\pm40^{\circ}$ with an efficiency
of $\sim 10^{-3}$ and at a distance of  $\pm10^{\circ}$ with an efficiency of $\sim 10^{-2}$. As a result, we use side lobe efficiency of $10^{-2}$ in our calculation. Details of side lobe BSA calibration can be found elsewhere (V. Samodurov et al. 2018, in preparation).

\begin{figure*}
\includegraphics[width=174mm,angle=-0]{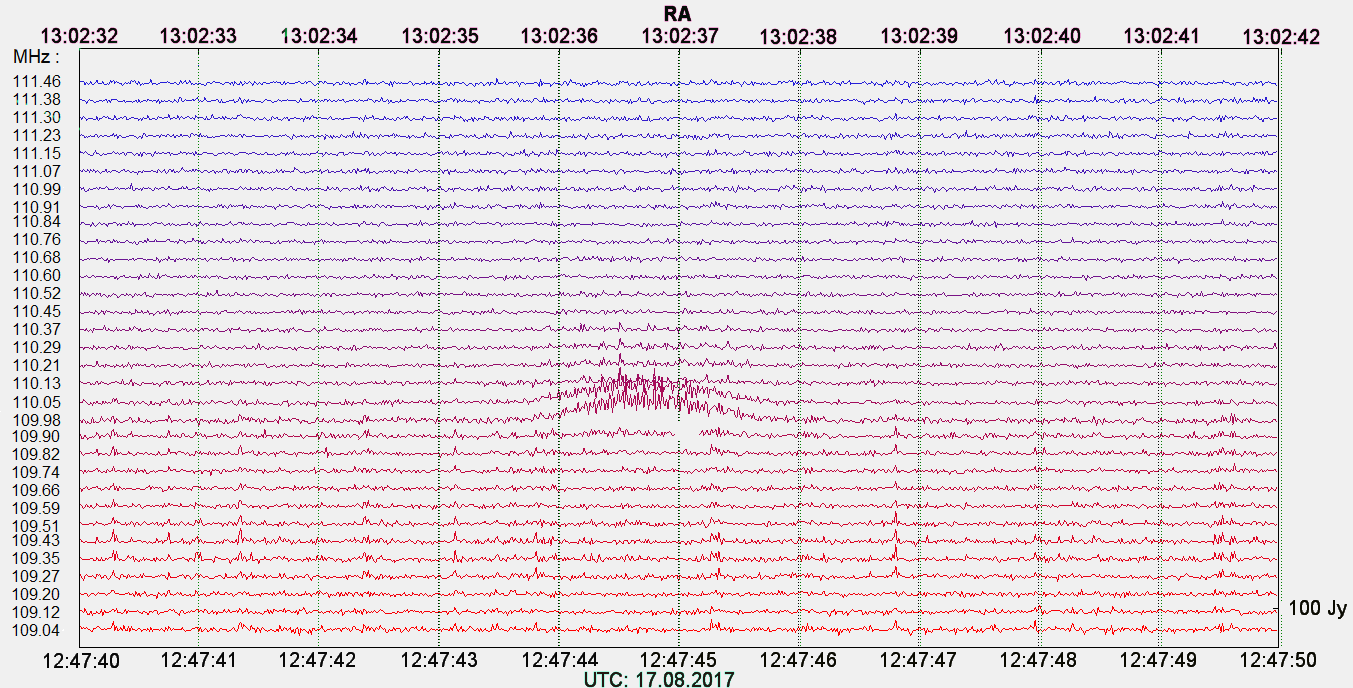}
\caption{BSA observation at 109 - 110.25 MHz frequency channels ($Y$-axis) around the GW170817 trigger time (lower $X$-axis). In particular, in the shown beam
toward decl. $(J2000) = -4.82 $ the nearly symmetric  signal is visible at frequency channels from 109.898 and up to  110.367 MHz.  The signal is visible in
the southern beams and centered at  (UTC) 12:47:44.7 i.e. 400 s later than the GW170817 trigger time. This is the most powerful signal of about 100 Jy  detected during 15 minutes since the GW170817 trigger.
It most likely has an artificial origin. }
\label{fig:BSA110}
\end{figure*}

We investigated all data sets around the GW170817 trigger recorded with BSA.
The only significant transient signal with a duration of 1.5 s centered at (UTC)
12:47:44.7 of about 100~Jy was detected in the southern beams
at 109.273-111.148 MHz (see Figure~\ref{fig:BSA110}). We mostly believe the signal has
a non-astrophysical nature because of the lack of a dispersion pattern, which is specific
for distant astrophysical objects such as, e.g., pulsars.  We can place a conservative upper limit of 30~Jy (it is equivalent to S/N$>10$) on a time scale of 10-60 s for
any astrophysical signal.

We can estimate the intensity of the possible astrophysical signal, which can be
detected by side lobes at the time of the GW170817 trigger. The FWHM of the side
lobe along R.A. is about 5 minutes. The OT
position was 2.5$\times$FWHM of the diagram at the trigger time of GW170817. Using a
projection of the BSA array toward the OT ($cos(78.65)=0.2$) and side lobe
efficiency $\sim 10^{-2}$, we estimate the upper limit on the transient radio
signal from the position of the OT as 15000~Jy. The value can be converted into the upper limit of a luminosity of the radio transient at the distance of the OT
$L^{\rm 110Mhz}_{(\rm iso)} < 5 \times 10^{40}$ erg s$^{-1}$ for a duration of the transient in the range of 10-60 s.

\section{SPI-ACS/INTEGRAL Data Analysis}
\label{sc:acs_app}

SPI-ACS experiment consisting of 91 BGO crystals is operating in a single energy channel with a lower threshold of $\sim$ 80 keV and time resolution of 50 ms \citep{kienlin_03}.
A script\footnote{\url{http://isdc.unige.ch/~savchenk/spiacs-online/spiacs-ipnlc.pl}} was used as the source of input SPI-ACS data.

GRB~170817A was detected in SPI-ACS at a 4.5$\sigma$ significance level. The burst onset is delayed to GW trigger to $\simeq2$~s. A third-order polynomial model was used to fit background signal
in time intervals $(-118, -30)$ s and $(300, 900)$~s (data before $-118$~s are not available). We found for GRB~170817A that there is a deviation of SPI-ACS count statistics from Poisson
distribution with a factor of $\simeq 1.26$ \citep[see, e.g., ][]{min10,min17}. Statistical uncertainties for SPI-ACS data were estimated, taking into account the factor of deviation.

We found the duration of the burst in SPI-ACS data, $T_{90}^{ACS} = 1.0 \pm 0.4$ s, to be significantly longer comparing with $T_{90} \simeq 0.1$ s derived in \cite{sav17}. Our value
is more consistent with GBM results obtained in our paper (see Figure~\ref{fig:acs-gbm} and Section~\ref{sc:GBM}) and in \cite{gol17}. The duration, $T_{90} = 1.0 \pm 0.4$ s,
characterizes the burst to be from a short (or type I) population with a probability of $\sim$ 74\% \citep[see Fig. 11 in ][]{min17}.

The burst fluence calculated for the (2, 3) s time interval since the GW trigger is $F = (1.8 \pm 0.4)\times10^{3}$ counts. The off-axis for the GRB~170817A source from the satellite pointing axis is 105 deg,
which is quite optimal for detection. In the paper by \cite{vig09}, it was shown that 1 SPI-ACS count corresponds on average to $\approx 10^{-10} $erg cm$^{-2}$ in the (75, 1000) keV range,
for directions orthogonal to the satellite pointing axis. Using the conversion factor, we derive $F = (1.8 \pm 0.4)\times10^{-7}$ erg cm$^{-2}$ in the (75, 1000) keV range for GRB~170817A.
This value is in agreement with one estimated in \cite{sav17} using a more complex method.

We did not find any precursor or extended emission components in SPI-ACS data at time scales from 0.05 up to 100 s in the time interval (-50, 200) s since the GW trigger and we estimated upper
limits on their intensity. At a time scale of 0.05 s, the upper limit on precursor activity is $S_{\rm prec} \simeq $ 250 counts or $S_{\rm prec} \simeq 2.5\times10^{-8} $ erg cm$^{-2}$ at the 3$\sigma$ significance level.
At a time scale of 50 s, the upper limit on extended emission activity is $S_{\rm EE} \simeq $ 8100 counts or $S_{\rm EE} \simeq 8.1\times10^{-7} $ erg cm$^{-2}$ at the 3$\sigma$ significance level.

Assuming redshift z = 0.00968 and luminosity distance $D_{L}$ = 42.0 Mpc for the source \citep{2009MNRAS.399..683J}, we estimated total isotropic energy release in the energy range (75, 1000) keV as
$E_{\rm iso}$ = $(3.9 \pm 0.9)\times10^{46}$ erg. One can estimate the upper limits on the precursor and extended emission component expressed as total isotropic energy release as well:
$E_{\rm iso}^{\rm prec}$ = $5.3\times10^{45}$ erg and $E_{\rm iso}^{\rm EE}$ = $1.7\times10^{47}$ erg, correspondingly.

\section{Magnetic Force-free Bomb}
\label{ap:bomb}

At the moment of the breakout, the jet, which was strongly  magnetically dominated close to the BH but became
mildly magnetically dominated during the confined propagation (Section \ref{spike}), starts to expand and accelerate.
As the pair density falls precipitously, the flow becomes more and more magnetically dominated. In this
appendix, we consider the explanation of a magnetic  force-free bomb -- radial motion of the post-breakout
highly magnetized plasma flows.

 Consider the radial expansion of force-free plasma carrying the toroidal \Bf. The relativistic force-free condition
 implies that the total EM force vanishes \citep{Gruzinov99}
 \be
 \E\; \nabla\cdot \E + {\bf J} \times \B=0 \rightarrow {\bf J}={\frac{({\bf E}\times{\bf B})\nabla\cdot{\bf E}+
({\bf B}\cdot\nabla\times{\bf B}-{\bf E}\cdot
\nabla\times{\bf E}){\bf B}}{B^2}},
 \ee \citep[for the GR
 formulation, see][]{2011MNRAS.418L..94K,2011PhRvD..83f4001L}.

 Separating the  angular and the  time-radial dependence,
 $B_\phi = B_\phi(t,r) b_\phi (\theta)$ and $ E_\theta
 = E_\theta(t,r) e_\theta (\theta)$, we find two types of solutions. The first is the solution self-similar in $Z=r/t$,
 \ba &&
 E_\theta(Z) = B_\phi(Z) =( 1-1/Z)B_0,
 \nn &&
 e_\theta (\theta) = b_\phi (\theta),
 \nn &&
 j_r=( 1-1/Z) \frac{ \partial_\theta (b_\phi \sin \theta) }{r \sin \theta},
 \nn &&
 \nabla\cdot \E =( 1-1/Z)\frac{\partial_\theta (b_\phi \sin \theta)}{r \sin \theta} = j_r.
 \label{eq:bomb}
 \ea
 Equation (\ref{eq:bomb}) represents solutions for the polar-angle-dependent expansion of force-free plasma into
 vacuum.  Poynting flux is zero at the edge of the outflow at $r=t$ and increases for smaller $r$. If the source is
 located at $r\ll t$, it is required that the source luminosity  increases with time $\propto t^2$ for the
 solution to be applicable.

 Second, looking for a solution of the type $B_\phi, E_\theta \propto 1/r$, we find that the following scaling satisfies the Maxwell equations
 \ba &&
 B_\phi (t,r) = E_\theta (t,r)  = \frac{ f(t-r)}{r},
 \nn  &&
  j_r= \frac{ f(t-r)}{r^2}  \frac{ \partial_\theta (b_\phi \sin \theta) }{r \sin \theta}
 \label{bom1}
 \ea
 where $f$ is an arbitrary function of the argument. This solution represents a force-free pulse  of initial shape $f(r)$, moving with the speed of light -- a force-free bomb.

  Both solutions (\ref{eq:bomb})  and  (\ref{bom1}) are different from EM fields in vacuum since there are non-zero currents and charge
  densities in the flow. A general time-dependent approach to force-free fields for the specific case of $\rho=j$ has been discussed by \cite{2014MNRAS.445.2500G}; see \cite{2011PhRvD..83l4035L} for a related time-dependent  solution in general relativity.

\end{document}